# Presolar Grains As Probes of Supernova Nucleosynthesis

Nan Liu[1*], Maria Lugaro[2,3,4,5], Jan Leitner[6,7*], Bradley S. Meyer[8], Maria Schönbächler[9]

[1]Institute for Astrophysical Research, Boston University, Boston, MA 02215, USA

*nanliu@bu.edu

*jan.leitner@mpic.de

[2]Konkoly Observatory, HUN-REN Research Centre for Astronomy and Earth Sciences, Konkoly Thege Miklós út 15-17., H-1121, Hungary

[3]CSFK, MTA Centre of Excellence, Budapest, Konkoly Thege Miklós út 15-17, H-1121, Hungary

[4]ELTE Eötvös Loránd University, Institute of Physics and Astronomy, Budapest 1117, Pázmány Péter sétány 1/A, Hungary

[5]School of Physics and Astronomy, Monash University, VIC 3800, Australia

[6]Max Planck Institute for Chemistry, D-55128 Mainz, Germany

[7]Institute of Earth Sciences, Heidelberg University, D-69120 Heidelberg, Germany

[8]Department of Physics and Astronomy, Clemson University, Clemson, SC 29634, USA

[9]Institute for Geochemistry and Petrology, ETH Zürich, 8092 Zurich, Switzerland

**ABSTRACT**

We provide an overview of the isotopic signatures of presolar supernova grains, specifically focusing on $^{44}$Ti-containing grains with robustly inferred supernova origins and their implications for nucleosynthesis and mixing mechanisms in supernovae. Recent technique advancements have enabled the differentiation between radiogenic (from $^{44}$Ti decay) and nonradiogenic $^{44}$Ca excesses in presolar grains, made possible by enhanced spatial resolution of Ca-Ti isotope analyses with the Cameca NanoSIMS (Nano-scale Secondary Ion Mass Spectrometer) instrument. Within the context of presolar supernova grain data, we discuss (*i*) the production of $^{44}$Ti in supernovae and the impact of interstellar medium heterogeneities on the galactic chemical evolution of $^{44}$Ca/$^{40}$Ca, (*ii*) the nucleosynthesis processes of neutron bursts and explosive H-burning in Type II supernovae, and (*iii*) challenges in identifying the progenitor supernovae for $^{54}$Cr-rich presolar nanospinel grains. Drawing on constraints and insights derived from presolar supernova grain data, we also provide an overview of our current understanding of the roles played by various supernova types – including Type II, Type Ia, and electron capture supernovae – in accounting for the diverse array of nucleosynthetic isotopic variations identified in bulk meteorites and meteoritic components. We briefly overview the potential mechanisms that have been proposed to explain these nucleosynthetic variations by describing the transport and distribution of presolar dust carriers in the protoplanetary disk. We highlight existing controversies in the interpretation of presolar grain data and meteoritic nucleosynthetic isotopic variations, while also outlining potential directions for future research.

# 1. INTRODUCTION

Supernovae, extraordinary cosmic events, are pivotal in the nucleosynthesis of the common elements that are essential for forming star-planet systems, including our solar system. In cosmochemistry, the term "supernova" frequently appears in literature discussions on nucleosynthetic isotope variations and the initial presence of short-lived nuclides, which have been identified in diverse meteoritic components and bulk meteorites (e.g., Leya et al. 2009; Burkhardt et al. 2011; Akram 2013; Cook & Schönbächler 2017; Ek et al. 2020; Schiller et al. 2020; Hopp et al. 2020; Nie et al. 2023). Some of these isotopic signatures have been suggested to originate from the heterogenous distribution of "supernova" nucleosynthesis products across the protoplanetary disk (e.g., Hutchison et al. 2022). "Supernova", in fact, is a broad term and encompasses various types, such as Type I, Type II, and electron-capture supernovae (ECSNe). Each type of supernovae contributes uniquely to element and isotope synthesis during galactic chemical evolution (GCE) – the process by which the abundances and isotopic compositions of the chemical elements in a galaxy, for example, the Milky Way Galaxy, change over time (Timmes et al. 1995; Prantzos et al. 2018; Kobayashi et al. 2020; Arcones & Thielemann 2023). Differentiating between nucleosynthesis products from the different supernova types and linking isotopic variations in meteoritic components to a specific supernova type are, therefore, crucial in understanding the origin and formation of the solar system.

Type II, Type Ib, and Type Ic supernovae, as classified spectroscopically, are all theoretically considered as core-collapse supernovae (CCSNe) originating from the collapse of the Fe core of massive stars (Modjaz et al. 2019; Burrows & Vartanyan 2021). The key similarity among them is the shared mechanism of the Fe-core collapse. The core collapse occurs when the central Fe core of a massive star, which is produced by previous burning phases (from He burning up to Si burning), cannot sustain gravity anymore via energy released by nuclear burning and undergoes a gravitational collapse, leading to the infalling matter to bounce back from the forming neutron star and, consequently, generating a cataclysmic explosion. Type II supernovae are commonly associated with red supergiants with extended H envelopes, Type Ib supernovae with progenitors such as Wolf-Rayet (WR) stars, where the outer layers of H were previously stripped by strong stellar winds. Type Ic supernovae are also associated with the WR stars, except that in this case the WR stars lost both their H and He layers. Compared to Type Ib and Type Ic supernovae, Type II supernovae are more common and thus provide a more significant contribution to the GCE (Kobayashi et al. 2020). Type II supernovae were historically believed to be the astrophysical sites for rapid neutron-capture process (*r*-process) that produces roughly half of the cosmic abundances of the elements that are heavier than Fe and up to Th and U (e.g., Hillebrandt et al. 1976; Meyer et al. 1992; Woosley & Hoffman 1992). Recent simulations and observations, however, have changed this view. Model simulations encountered difficulties in producing, in typical Type II supernovae, the necessary neutron-rich environment for the *r*-process (e.g., Fischer et al. 2010). In addition, several pieces of evidence imply the infrequent occurrence of *r*-process events and challenge the suggestion that the frequently occurring Type II supernovae are the primary sites for the *r*-process, including large scatter in *r*-process elements detected among metal-poor stars (Sneden et al. 2008) and the initial $^{244}$Pu abundances inferred for the early solar system and the present-day interstellar medium (ISM) based on meteorite and deep-sea measurements, respectively (Hotokezaka et al. 2015). The current consensus posits that the *r*-process could occur in rare Type II supernovae such as magnetorotational supernovae, alongside the mergers of compact objects such as neutron stars (NS) and black holes (Thielemann et al. 2017). Indeed, the gravitational wave event GW170817 that was associated to a short γ-ray burst and caused by a NS-NS merger event, was observed as a kilonova *r*-process event (Kilpatrick et al. 2017; Pian et al. 2017).

Type Ia supernovae instead originate from the thermonuclear explosion of white dwarf stars (see Ruiter 2020 for a review). In a binary star system, where two stars orbit each other, if one or both of the stars are white dwarfs, they can accrete mass from the companion or merge, eventually undergoing a cataclysmic event and leading to a Type Ia supernova. Type Ia supernovae, alongside Type II supernovae, contribute significantly to the Fe-group elements present in the ISM but at a later time compared to Type II supernovae (Tinsley 1979; Greggio & Renzini 1983). This temporal variation in the contributions of Type II and Type Ia supernovae to the galactic Fe-group element inventory is caused by the longer evolutionary time of Type Ia supernovae, given the time it takes for its progenitor stars to transition to the white dwarf phase and the following accretion and merger.

Electron capture supernovae represent another intriguing class, which is triggered by electron capture in the core and could occur as thermonuclear or collapsing ECSNe (Jones et al. 2019b). These events were predicted to occur in 8−10 $M_\odot$ stars with degenerate O/Ne/Mg cores (Nomoto 1984, 1987). Model predictions of the stellar evolution within this mass regime, however, remain uncertain (Jones et al. 2013), and observational evidence of their existence was lacking. Thus,

ECSNe had long faced skepticism until a recent breakthrough detection confirmed the occurrence of an ECSN (Hiramatsu et al. 2021). Wanajo et al. (2011) suggested the potential for a weak *r*-process, i.e., reaching only up to atomic mass ≈ 100, in ECSNe, albeit with some debate surrounding this proposal (Jones et al. 2016).

Accurate nucleosynthesis model predictions for all these supernova types are critical for distinguishing between their nucleosynthesis products and linking isotopic variations in meteoritic components to the specific supernova type. Many challenges persist due to uncertainties in the structure and evolution of the progenitor star, the explosion hydrodynamics, and the nuclear reaction rates, as well as neutrino transport related to the explosion mechanism of Type II supernovae, the still debated evolution channel(s) for Type Ia supernovae, and the unknown likelihood of 8–10 $M_\odot$ stars evolving to ECSNe. Direct observations of isotope abundances within these supernovae offer ideal tests for nucleosynthesis models. Such measurements, however, are currently limited to a few radioactive isotopes (*e.g.*, $^{56}$Ni and $^{44}$Ti; Diehl 2017).

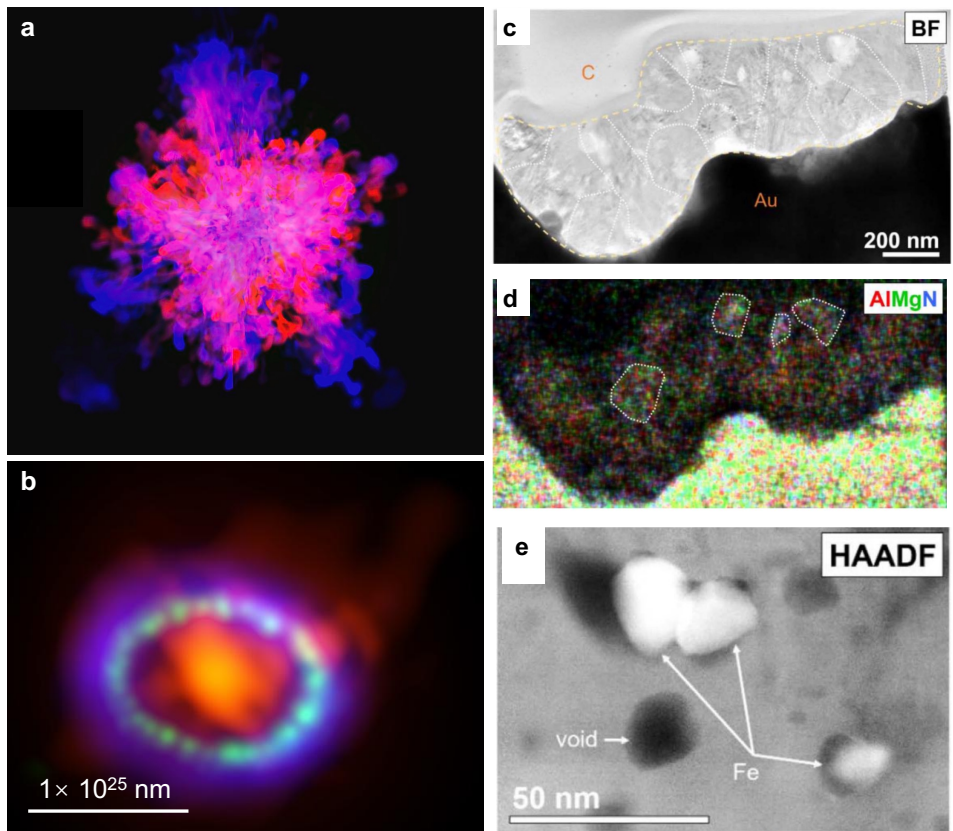

**Figure 1**. *Panel (a) presents the 3D simulation of a CCSN at 333 seconds post-ignition with elements represented in different colors: Ni (blue), O (red), C (green), and Ni-O mixture (pink), calculated by Dr. H.-T. Janka (the Max Planck Institute for Astrophysics). Panel (b) shows the*

*Atacama Large Millimeter/submillimeter Array (ALMA) telescope multiwavelength observation of SN 1987A (Indebetouw et al. 2014), highlighting dust (red), the circumstellar gas ring (green), and the supernova blast-induced shock wave (purple) (image credit: Alexandra Angelich). Panels (c) through (e) feature the bright field (BF) image, energy-dispersive X-ray composite image, and high-angle annular dark field (HAADF) image, respectively, of a Type-X presolar SiC grain, which formed in the remnant of a Type II supernova explosion that occurred before the solar system formation (Singerling et al. 2021). The X grain is denoted by the contour line in panel (c), in which the letters "c" and "Au" denote carbon deposited on top of the grain during sample preparation and the gold foil on which the grain was initially dispersed, respectively (see Singerling et al. 2021 for details). In panel (d), the high Mg concentration (in green) of the X grain results from the radiogenic decay of $^{26}Al$ ($t_{1/2}$ =0.72 Ma), since X grains are enriched in Al with high initial $^{26}Al/^{27}Al$ ratios (Hoppe et al. 2023; Liu et al. 2024a)*

In the aftermath of a supernova explosion, the condensation of dust becomes plausible when the temperature undergoes a significant drop (below ~2500 K) while maintaining a high density in the ejecta (Sarangi et al. 2018). Current spectroscopic observations suggest that (*i*) dust forms efficiently in Type II supernova remnants (Wesson & Bevan 2021), e.g., SN 1987A (Fig. 1b), (*ii*) Type Ia supernovae show no definitive evidence of dust, potentially due to low formation efficiencies and/or elevated destruction rates (Douvion et al. 2001), and (*iii*) dust formation in ECSNe remains unknown because of the only recent first detection of such an event (Hiramatsu et al. 2021). Considering the observed efficient dust formation around and high occurrence frequencies of Type II supernovae, dust from supernovae in the ISM is anticipated to come predominantly from Type II supernovae. Supernova dust is expected to have served as one of the foundational constituents of the ISM material that formed the solar system at about 4.6 Ga ago and then became incorporated and preserved in primitive extraterrestrial materials – the leftover building blocks of the planets.

Consistent with this expectation, ancient stardust grains, which include contributions predominantly from asymptotic giant branch (AGB) stars, Type II supernovae, and possibly novae, have been identified in primitive extraterrestrial materials, offering a direct probe into the conditions for the birth of the solar system (Nittler & Ciesla 2016; Liu 2024). These grains range from submicron up to tens of microns in size and are diverse in chemical composition, including phases such as silicon carbide (SiC), silicon nitride ($Si_3N_4$), graphite, oxides, and silicates. Certain nanospinel grains, showing large $^{54}$Cr excesses, have been suggested to originate from Type Ia supernovae or ECSNe (Nittler et al. 2018), albeit a potential Type II supernova origin was also proposed (Dauphas et al. 2010; Qin et al. 2011; den Hartogh et al. 2022). Collectively termed as "presolar grains", these minute cosmic fossils recorded the nucleosynthetic legacy of their parent stars. In the literature, presolar SiC and O-rich grains are divided into different groups based on their isotopic compositions, and presolar graphite grains are divided into high-density and low-density grains based on their density separation during the acid extraction procedure (Amari et al. 1994; Stephan et al. 2024). The adopted classification schemes for various presolar phases are useful in identifying their stellar origins (see Liu 2024 for a recent review).

Figure 1 exemplifies the interconnection between the laboratory studies of these grains and the broader astronomical and astrophysical research in understanding Type II supernovae. Meteorite investigations point to the role of Type II supernovae as prolific dust factories (Hoppe et al. 2000; Nittler & Ciesla 2016; Liu 2024), in line with the implication from the multiwavelength observations of SN 1987A remnant via ALMA (Fig. 1b; Indebetouw et al. 2014). X grains are a group of presolar SiC grains with robust Type II supernova origins inferred from their multielement isotope data (Nittler et al. 1996; Lin et al. 2010; Xu et al. 2015; Liu et al. 2018a, 2024a). The presence of Fe subgrains within an X grain shown in Fig. 1e offers a granular view of the mixing process that ensues following a supernova explosion (Singerling et al. 2021), phenomena that can now be juxtaposed with three-dimensional (3D) simulations, as exemplified in Fig. 1a (Wongwathanarat et al. 2015, 2017). By piecing together insights from large-scale computational models, telescope observations, and microscopic-scale meteoritic data, we can significantly enhance our capability to comprehend the evolution of supernovae, their explosive nucleosynthesis, and dust formation in these spectacular celestial events.

This review will delve into recent advancements in expanding the inventory and multielement isotope data of presolar supernova grains identified within primitive meteorites, and the interpretation of such data, aimed at unraveling new implications for supernova nucleosynthesis. Guided by presolar supernova grain data, we will focus on a comparative analysis of the stellar nucleosynthesis of the three supernova types — Type II, Type Ia, and ECSNe. Finally, we will contextualize these implications within the realm of nucleosynthetic isotope variations in bulk meteorites. This comprehensive examination will guide our future exploration of (*i*) the supernova type that is responsible for the observed nucleosynthetic variations in meteorites and (*ii*) the early solar system processes that led to the widespread distribution of isotopic variations across the protoplanetary disk, from the perspective of existing presolar grain data.

## 2. PRESOLAR SUPERNOVA GRAINS AND THEIR ISOTOPIC SIGNATURES

### 2.1. Titanium-44: A Smoking Gun of Supernova Nucleosynthesis

#### 2.1.1. Production of $^{44}Ti$ in Supernovae

In the context of supernova nucleosynthesis, the radioactive isotope $^{44}$Ti ($t_{1/2} = 60$ a) emerges as a compelling indicator, particularly for Type II supernovae. This unequivocal status stems from two key observations: (*i*) $^{44}$Ti is predicted to be produced primarily during explosive O burning and $\alpha$-rich freeze outs from nuclear statistical equilibrium (NSE) in the deep layers of Type II supernovae (The et al. 2006; Sieverding et al. 2023), and (*ii*) unambiguous detection of $\gamma$-ray emissions from the decay of $^{44}$Ti was reported in the Type II supernova remnants SN 1987a and Cassiopeia A (Cas A; Grefenstette et al. 2014, 2017; Boggs et al. 2015). In contrast, Type Ia supernovae and ECSNe are anticipated to yield $^{44}$Ti in quantities one order of magnitude lower than Type II supernovae (Nomoto 1984,1987,1997; The et al. 2006). This is because, in the former two types of supernovae, $^{44}$Ti is produced only through explosive He burning since explosive O burning and the conditions for $\alpha$-rich freeze outs are typically not attained (Woosley et al. 1986). Notably, the detection of $\gamma$-ray emissions from $^{44}$Ti in Type Ia supernovae remains elusive (Weinberger et al. 2020), and no detection of $^{44}$Ti $\gamma$-ray emissions in ECSNe has been reported. *Integrating these pieces of evidence suggests that the initial presence of $^{44}Ti$ in a presolar grain provides a direct piece of evidence to support its origin in supernovae, although further examination based on other isotope signatures is always required to pinpoint the specific supernova type.*

The absence of initial $^{44}$Ti in a presolar grain, however, does not conclusively exclude a supernova origin, considering its low productions in Type Ia supernovae and ECSNe and the potential lack of $^{44}$Ti (despite its production in the deep region) incorporated into the parent Type II supernova ejecta of the grain. Nevertheless, here we focus our discussions primarily on presolar grains with inferred initial presence of $^{44}$Ti because of their robustly determined supernova origins.

### 2.1.2. Detection of $^{44}Ti$ in Presolar Grains

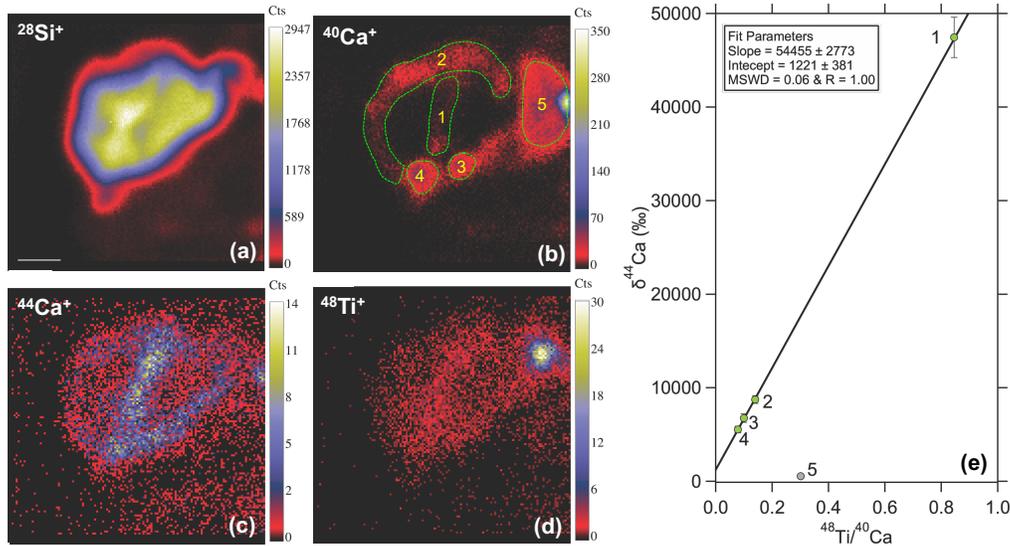

**Figure 2.** *NanoSIMS Ca-Ti isotope data of a Type-X SiC grain from Liu et al. (2023a). Panels (a) to (d) show the $^{28}Si$, $^{40}Ca$, $^{44}Ca$, and $^{48}Ti$ ion images, respectively, collected from the grain at a spatial resolution of ~150 nm. The white scalebar in panel (a) denotes 500 nm. In panel (b), five different regions of interest (ROIs) were defined for reducing the Ca-Ti isotope data shown in panel (e), where $\delta^{44}Ca$ was calculated as $[(^{44}Ca/^{40}Ca)_{grain} / (^{44}Ca/^{40}Ca)_{standard} - 1] \times 1000$, and the linear fit was obtained using the CEREsFit.xlsm tool of Stephan & Trappitsch (2023). MSWD stands for mean squared weighted deviation, and R is the Pearson correlation coefficient. Errors are all $1\sigma$. The data for ROI #5 was excluded from the linear fit because of its significant deviation. This deviation is likely predominantly due to some Ti contaminant at the grain edge, resulting in an increase in the $^{48}Ti/^{40}Ca$ ratio.*

Due to the short half-life of $^{44}Ti$ compared to the residency of presolar grains in the ISM (up to ~1 Ga; Gail et al. 2009; Heck et al. 2020) and within the solar system (4.6 Ga), any $^{44}Ti$ atoms initially present within a presolar grain have already fully decayed to $^{44}Ca$. This decay process can potentially yield a detectable $^{44}Ca/^{40}Ca$ excess in the grain, depending on its Ca/Ti ratio (i.e., $^{40}Ca/^{48}Ti$ in Fig. 2e). The analysis of Ca-Ti isotope systematics in presolar grains is routinely conducted using the Cameca Nanoscale Secondary Ion Mass Spectrometer (NanoSIMS; Hoppe 2006). In the past, direct, quantitative distinction between radiogenic (from $^{44}Ti$ decay) and nonradiogenic $^{44}Ca$ excess was challenging, leading to the derivation of the initial $^{44}Ti/^{48}Ti$ ratio based on the assumption that the detected $^{44}Ca$ excess in a presolar grain is fully due to the $^{44}Ti$ decay (Nittler et al. 1996). Recent advancements in spatial resolution, enabled by the Hyperion radio-frequency plasma $O^-$ ion source manufactured by Oregon Physics (Malherbe et al. 2016), have empowered NanoSIMS analyses to directly discriminate between radiogenic and nonradiogenic $^{44}Ca$ excess for µm-sized and/or Ti-rich (and Ca-poor) presolar grains (Liu et al. 2023b; Nittler et al. 2023; Leitner et al. 2024).

Figure 2 provides an illustrative example of high-resolution Ca and Ti imaging data. The observed correlation of Ca and Ti ions at high spatial resolution enables the differentiation between radiogenic and nonradiogenic $^{44}Ca$ excess in a Type-X SiC grain. The evident correlation between the distributions of $^{44}Ca^+$ and $^{48}Ti^+$ (rather than $^{40}Ca^+$) within the grain directly implies that $^{44}Ca^+$ was initially present in the grain as $^{44}Ti^+$. In other words, the predominant radiogenic origin of $^{44}Ca^+$ as the decay product of $^{44}Ti^+$ is demonstrated by the fact that $^{44}Ca^+$ is not spatially correlated with $^{40}Ca^+$ that shares the same chemical property with $^{44}Ca^+$ but instead with $^{48}Ti^+$ that is chemically the same as $^{44}Ti^+$. The variable $^{48}Ti/^{40}Ca$ ratios across the grain further allow for the derivation of the initial $^{44}Ti/^{48}Ti$ ratio based on the linear fit, known as an internal isochron, shown in Fig. 2(e). The slope of the isochron yields an initial $^{44}Ti/^{48}Ti$ ratio of $1.18 \pm 0.06$, and the y-axis intercept corresponds to a nonradiogenic $^{44}Ca$ excess of $1221 \pm 381$ ‰, revealing the production of both $^{44}Ca$ and $^{44}Ti$ during Type II supernova nucleosynthesis. Thus, high-resolution NanoSIMS imaging of Ca and Ti isotopes provides a powerful tool for evaluating the origin of $^{44}Ca$ excesses in presolar grains, thereby allowing for a more accurate determination of both the initial $^{44}Ti/^{48}Ti$ and nonradiogenic $^{44}Ca/^{40}Ca$ ratios in these grains.

To date, $^{44}Ca$ excesses[1] have been detected in various presolar phases and groups, including a substantial fraction of Type X SiC grains (Nittler et al. 1996; Lin et al. 2010; Xu et al. 2015), a few Type C and ungrouped SiC grains (Hoppe et al. 2010; Xu et al. 2015), a single Type AB SiC grain (high-resolution imaging; Liu et al. 2023b), certain presolar graphite grains (Nittler et al. 1996; Amari et al. 2014), a few Group 1 and Group 4 silicates (Leitner & Hoppe 2022; Leitner et al. 2024), a few Group 4 oxides (Nittler et al. 2008), one Group 3 oxide (Nittler et al. 2011), and an ungrouped $^{16}O$-rich oxide (Gyngard et al. 2010). Among the phases analyzed, SiC is one of the most suitable phases for inferring the initial $^{44}Ti/^{48}Ti$ ratio because of its Ti enrichment, low intrinsic Ca consent, and typically large size. In contrast, oxides and silicates[2] typically show higher Ca concentrations compared to Ti, resulting in their overall smaller radiogenic $^{44}$Ca excesses (e.g., Nittler et al. 2008). The presence of intrinsic Ca in these phases, coupled with small $^{44}$Ca excesses, introduce ambiguities in the interpretation of their $^{44}$Ca excesses. This is because (*i*) both $^{44}$Ca and $^{44}$Ti can be produced by stellar nucleosynthesis in a star, and, additionally, (*ii*) the initial composition of the star may differ from the solar composition because of the effects of GCE and heterogeneities in the ISM. Heterogenous ISM compositions could mean local enrichments of $^{44}$Ti and/or $^{44}$Ca in the ISM (The et al. 2006), and stars that subsequently form in such zones may exhibit $^{44}$Ca excesses in their initial composition relative to the broader Galactic baseline at the time. This results in their stardust grains to display GCE-related nonradiogenic $^{44}$Ca excesses.

---

[1] The $^{44}Ca$ excesses detected in presolar grains so far have mostly been collected using NanoSIMS instruments equipped with the previous generation of $O^-$ plasma ion source at poorer spatial resolutions, compared to the data shown in Fig. 2. These previous studies, in most cases, could not quantitively distinguish between radiogenic and nonradiogenic contributions to their observed $^{44}Ca$ excesses and assumed that the observed $^{44}Ca$ excesses were solely due to $^{44}Ti$ decay.

Finally, the determination of the initial $^{44}$Ti/$^{48}$Ti ratio from SIMS Ca-Ti isotope data of presolar SiC and potentially graphite grains, faces additional uncertainties related to the SIMS relative sensitivity factor (RSF), which is needed for determining the Ca/Ti elemental ratio. The RSF, which corrects for different ionization efficiencies of different elements in the SIMS instrument, can be different for different materials and must be determined through measurements of standards. Currently, this factor is calibrated using NIST O-rich glass standards due to the absence of viable alternatives. Recent findings by Liu et al. (2024) reported that the utilization of O-rich standards for calibrating the SIMS RSF value for the elemental Mg/Al ratio results in a factor of two reduction in the derived initial $^{26}$Al/$^{27}$Al ratio for presolar SiC grains. This is caused by different sample versus standard oxidation states. This discrepancy may have similarly impacted the inferred initial $^{44}$Ti/$^{48}$Ti ratios of presolar SiC and graphite grains. Consequently, it is essential to bear in mind a factor of two potential, systematic uncertainty when comparing presolar SiC and graphite grain data to model calculations.

*In summary, high-resolution NanoSIMS imaging has enabled quantitative distinction between nonradiogenic and radiogenic $^{44}$Ca excesses in μm-sized presolar grains. The inferred initial $^{44}$Ti/$^{48}$Ti ratios of presolar SiC and graphite grains, however, still suffer from uncertainties in the SIMS RSF value, which needs to be addressed in future work.*

---

[2]In addition, presolar silicates are measured *in situ* in meteoritic matrices, which leads to unavoidably sampling adjacent Ca-rich matrix materials if present and, in turn, diluted isotopic anomalies.

### 2.1.3. The Initial Presence of $^{44}$Ti in Type-X SiC Grains

2.1.3.1. Hypothesis: Production of $^{44}$Ti in rare Type II supernovae

The production of $^{44}$Ti is recognized as one of the contributing sources of the cosmic abundance of $^{44}$Ca. It is, however, debated whether typical Type II supernovae alone govern the production of $^{44}$Ti in the Galaxy. Two key observations challenge this view (The et al. 2006): (*i*) given the short half-life of $^{44}$Ti and the Galactic Type II supernova occurrence rate of approximately three per 100 a, several $^{44}$Ti γ-ray sources are expected, yet only one, the 360-year-old Cas A, was definitively identified in the Galaxy; and (*ii*) Cas A is located far from the inner Galaxy, contrary to the expectation that the inner region, rich in young massive stars, should be the primary location for $^{44}$Ti γ-ray sources if typical Type II supernovae are the main contributors of $^{44}$Ti. In light of these inconsistencies, The et al. (2006) proposed that either (*i*) Type II supernovae have been improbably rare in the Galaxy during the past few centuries, or (*ii*) $^{44}$Ti is not predominantly produced by typical supernovae but by rare events.

Therefore, the question arises whether the lack of inferred initial $^{44}$Ti in some X grains may provide evidence for the idea that $^{44}$Ti is not produced in all Type II supernovae or reflects reduced incorporation of $^{44}$Ti-containing supernova materials into some of the grains. X grains are characterized by large $^{28}$Si excesses (Nittler et al. 1996), likely due to explosive O burning in the Si/S zone (see Fig. 3) (Lin et al. 2010) and/or α-rich freeze outs in the Fe/Ni zone (see Fig. 3) (Hoppe et al. 2000; Liu et al. 2024a). Explosive O burning in the Si/S zone significantly overproduces $^{28}$Si, while α-rich freeze outs in the Fe/Ni zone can overproduce both $^{28}$Si and $^{29}$Si with $^{29}$Si being produced by proton capture onto $^{28}$Si because of the free protons released by neutrino-nucleus reactions there (Liu et al. 2024a). The latter explains the fact that, in the plot of $\delta^{29}$Si versus $\delta^{30}$Si, X grains predominantly fall along a slope-2/3 line instead of the slope-1 line that is predicted if the Si/S zone with pure $^{28}$Si dominantly controls the $^{28}$Si excesses of X grains (Fig. 4a). In a Type II supernova, predominant production of $^{44}$Ti is predicted by α-rich freeze outs and explosive O burning (The et al. 1998). This leads to the expectation that if the parent supernova of an X grain produced $^{44}$Ti in these zones, $^{28}$Si and $^{44}$Ti could be incorporated into the grain together. In turn, the absence of inferred initial $^{44}$Ti in some X grains may imply the lack of $^{44}$Ti production in their parent supernovae. Let us examine this scenario below in detail based on X grain data compiled in the Presolar Grain Database (PGD; Stephan et al. 2024).

### 2.1.3.2. X grain formation: Large-scale versus local mixing scenarios

To compare the isotopic composition of an X grain with nucleosynthesis models for Type II supernova explosions, parametrized mixing of materials ejected from varying zones of the parent star has been traditionally invoked (Hoppe et al. 1996b; Nittler et al. 1996). This is because mixing processes in state-of-the-art 3D supernova simulations are quite uncertain and not well constrained (Müller 2020). Thus, comparing unmixed one-dimensional (1D) supernova nucleosynthesis model predictions for the different shells shown in Fig. 3 with the multielement isotopic compositions of X grains, offers a feasible approach for constraining the involved nucleosynthesis processes (Lin et al. 2010).

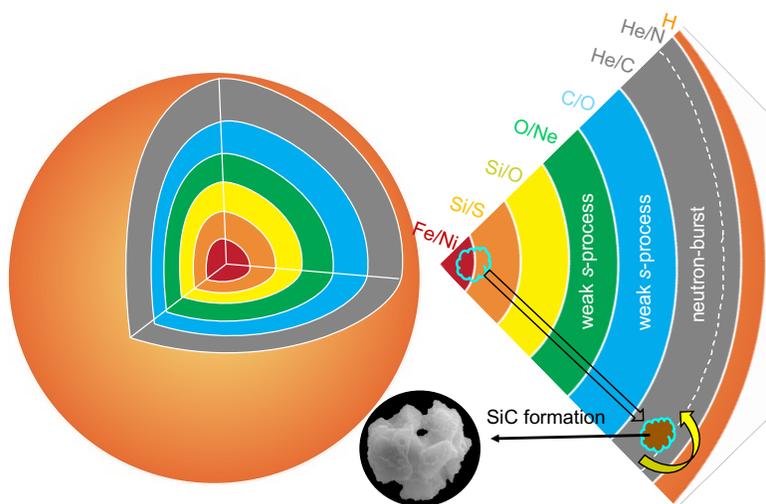

**Figure 3**. *Illustration of the onion shell structure of a presupernova massive star. Each zone (not to scale) is labeled by the two most abundant elements it contains (Meyer et al. 1995). On the right, illustrated is the formation process of X grains after the explosion when assuming that the grains incorporated material from the Fe/Ni zone adjacent to the central neutron star, the inner Si/S zone, and the He (and possibly H) shells.*

Traditionally, it was proposed that X grains incorporated materials predominantly from the Si/S and He/C zones (but not the shells inbetween), with the former contributing $^{28}$Si-rich and the latter contributing C-rich material (C>O) essential for the formation of SiC grains (Amari et al. 1992; Nittler et al. 1996). Additional evidence from Ca-Ti isotope systematics further implicated the mixing of Fe/Ni material in forming X grains (Hoppe et al. 2000; Lin et al. 2010), although the Fe isotopic compositions of X grains seem to suggest otherwise (Marhas et al. 2008). This is because, as shown by Marhas et al. (2008), the 25 $M_\odot$ CCSN model of Rauscher et al. (2002) predicts large $^{54}$Fe and $^{56}$Fe enrichments in the Fe/Ni zone, which are not observed in any of the X grains studied so far. According to Woosley and Hoffman (1992), whether α-rich freeze outs in the Fe/Ni zone predominantly produce $^{56}$Fe (the decay product of $^{56}$Ni) or $^{54}$Fe depends on the neutron richness there. Since the model predictions for $^{54}$Fe/$^{56}$Fe strongly depend on the neutron-rich condition during alpha-rich freeze outs, which is poorly constrained, more work is needed to interpret the Fe isotope data. More importantly, Fe contamination is a common problem for presolar SiC measurements (e.g., Trappitsch et al. 2018). NanoSIMS Fe isotope analyses of X grains at improved spatial resolution will shed further light onto the mixing process in their parent supernovae.

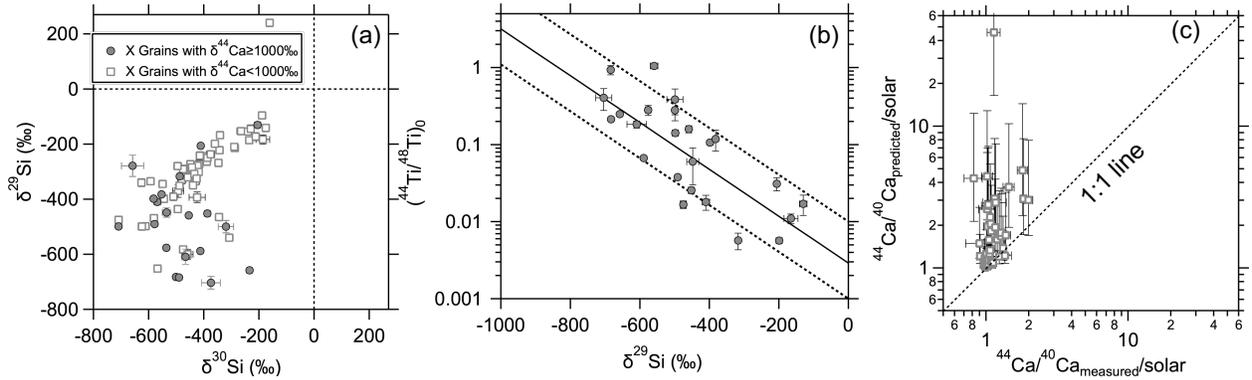

**Figure 4.** *Panel (a) shows Si isotope data for 26 SiC X grains with $^{44}$Ca excesses ⪞ 1000 ‰ (Amari et al. 1992, 2003; Nittler et al. 1996; Besmehn & Hoppe 2003; Lin et al. 2010; Xu et al. 2015; Liu et al. 2023a) and 53 SiC X grains from Lin et al. (2010) and Liu et al. (2018a, 2023a) with smaller $^{44}$Ca excesses (< 500 ‰) and available Ca and Ti concentration data. Panel (b) illustrates isotope data for the 26 SiC X grains with large $^{44}$Ca excesses. The solid line is an exponential fit to the grain data, and 90% of the grains lie within the dashed lines. Panel (c) shows predicted versus measured values of the 53 SiC X grains with smaller $^{44}$Ca excesses. See the main text for the calculation of $^{44}$Ca/$^{40}$Ca$_{predicted}$/solar.*

Challenging the conventional notion of large-scale, selective mixing, Pignatari et al. (2013b) posited that X grains could derive solely from the He shell and nearby regions, attributing their $^{28}$Si enrichments to high-temperature α-capture in a newly identified zone, the so-called C/Si zone, produced in a higher-than-standard energetic 1D supernova model. The energy required for forming the C/Si zone in the model of Pignatari et al. (2013b) corresponds to hypernovae[3], which are more massive and rarer than typical CCSNe. Several subsequent studies supported this local mixing theory to explain the multielement isotopic compositions of X grains (Xu et al. 2015; Hoppe et al. 2018, 2023, 2024; Pignatari et al. 2018; den Hartogh et al. 2022; Schofield et al. 2022). However, these data-model comparisons relied on an ad hoc mixing calculation method with intrinsic uncertainties because of its random selection and weighting of different shell materials.

---

[3]A hypernova is an extremely energetic type of supernovae resulting from the Fe-core collapse in massive stars (>30 $M_\odot$), leading to the formation of black holes. It ejects material with exceptionally high kinetic energy, around ten times greater than typical supernovae (Moriya et al. 2018). Alternatively, if asymmetric explosions are typical of CCSNe, such an energetic explosive condition may be achieved in certain directions during a typical CCSN explosion.

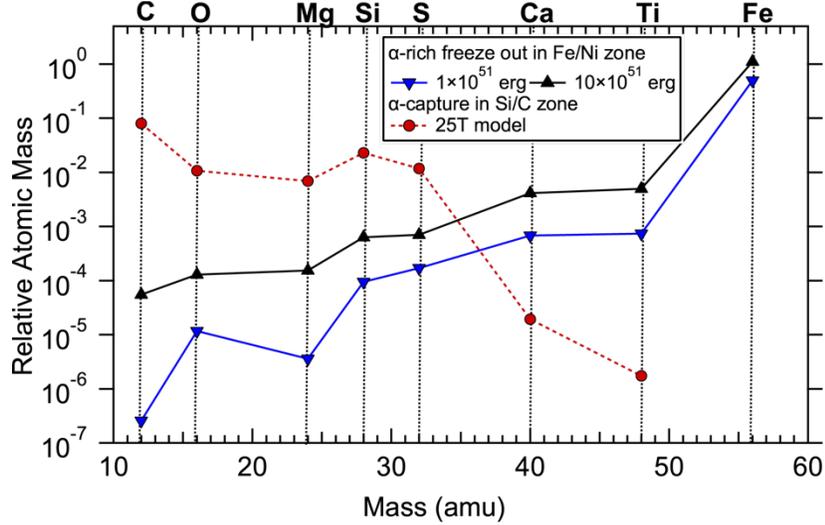

**Figure 5.** *Plot comparing the productions of α elements by (i) α-capture in the C/Si zone of the 25T CCSN model of Pignatari et al. (2013b) (in red) and by α-rich freeze outs in the 25 $M_\odot$ model from Liu et al. (2024a) at two different explosion energies (in black and blue). These latter calculations (Rauscher-Meyer models hereafter) explored explosive nucleosynthesis based on the 25 $M_\odot$ presupernova stellar model of Rauscher et al. (2002) and the explosion model and reaction network of Bojazi & Meyer (2014). Iron-56 is the decay product of $^{56}Ni$ made by NSE and the most abundant Fe isotope in this region. Nickel-56 decays to $^{56}Co$ with $t_{1/2}$= 6.1 d, which then decays to $^{56}Fe$ with $t_{1/2}$ = 77 d.*

Addressing these uncertainties, several recent investigations have reported correlations between different isotope ratios to deconvolute the isotopic signatures of different contributing supernova zones, thereby enhancing the accuracy in constraining supernova nucleosynthesis (Liu et al. 2018a, 2024a). For example, Liu et al. (2024a) highlighted a shortfall of the local mixing hypothesis: although Hoppe et al. (2023) reproduced their X grain isotopic compositions by invoking the C/Si zone and explosive H burning (i.e., the localized mixing scenario) based on the 1D, parametric models of Pignatari et al. (2015), the explosive H-burning zone ($^{30}Si$-rich) produces much higher $^{26}Al/^{27}Al$ ratios than the C/Si zone ($^{28}Si$-rich), which consequently implies a positive trend between $^{26}$Al/$^{27}$Al and $\delta^{30}$Si. In contrast to this expectation, a negative trend was observed among X grains (Liu et al. 2024a). The observed positive correlation between $^{26}$Al and $^{28}$Si excesses in X grains may instead support a large-scale mixing scenario (Fig. 3), in which neutrino-nucleus interactions within the Fe/Ni zone foster a proton-rich condition that enables substantial $^{26}$Al production, alongside $^{28}$Si synthesis via $\alpha$-rich freeze outs in the same region. More detailed calculations of neutrino-nucleus reactions following a supernova explosion are currently being pursued to investigate modelling uncertainties, based on which constraints can be derived on relevant parameters based on the X grain data (Walls & Meyer 2024).

Another example is related to the abundance pattern of $\alpha$ elements. The $\alpha$-rich freeze outs in the Fe/Ni zone and $\alpha$-capture in the C/Si zone can produce similar isotopic signatures (Liu et al. 2018a; Schofield et al. 2022), e.g., enrichments in $^{44}$Ti, $^{40}$Ca, $^{48}$Ti. Current models, however, predict that the two processes lead to distinct abundance patterns for $\alpha$ elements for the following reasons (Fig. 5). Alpha-rich freeze outs can modify the NSE products that are characterized by a peak at $^{56}$Ni because of its highest binding energy (Woosley & Hoffman 1992). In comparison, the seeds for $\alpha$-capture in the C/Si zone are the $^{12}$C produced by He burning. While $\alpha$-capture onto $^{12}$C at high temperatures can lead to efficient production of heavier $\alpha$ elements such as Si and S, the efficiency of the $\alpha$-capture, however, drops significantly in synthesizing heavier elements in the C/Si zone.

The isotopic compositions of X grains argue against the low Ti/Si and Ti/Ca ratios that characterize $\alpha$-capture in the C/Si zone in the models of Pignatari et al. (2013b, 2015) and thus prefer the large-scale mixing scenario. The negative correlation between the radiogenic $^{49}$Ti (from $^{49}$V decay) and $^{28}$Si excesses of X grains (Liu et al. 2018a, 2023a) can be reproduced by mixing between materials from the Si/S and He/C zones (Liu et al. 2023a). In addition, with respect to $^{49,50}$Ti excesses in X grains, much higher than expected $^{46}$Ca excesses have been observed in Si$_3$N$_4$ grains (Liu et al. 2024b), which likely condensed out of the same supernova ejecta with X grains but at lower temperatures according to their isotopic similarities (Nittler et al. 1995; Hoppe et al. 1996a; Lin et al. 2010). This observation can be explained by mixing materials from both the Fe/Ni zone and the He/C zone because $^{48}$Ti is produced in a larger quantity than $^{40}$Ca in the Fe/Ni zone so that its mixing with the He/C zone leads to larger reduction in $^{49,50}$Ti/$^{48}$Ti than in $^{46}$Ca/$^{40}$Ca in the ejecta (Liu et al. 2024b). Thus, both the observations argue against the low Ti/Si and Ti/Ca ratios that characterize $\alpha$-capture in the C/Si zone (Fig. 5) in the models of Pignatari et al. (2013b,

2015).

Further information can be derived from current 3D CCSN explosion model simulations and astronomical observations. Observations of CCSN remnants show that $^{56}$Ni and $^{44}$Ti knots from the innermost region lie outside of the central region of the remnants, supporting large-scale macroscopic mixing (Grefenstette et al. 2017). The microscopic mixing required to form dust with material from the inner and the outer layers (as required in the large-scale mixing scenario) is supported by some phenomenon observed in 3D hydrodynamic simulations. The 3D models predict that the outward-expanding shock can become decelerated due to a Rayleigh–Taylor-unstable density inversion near the He/H interface (Hammer et al. 2010; Wongwathanarat et al. 2017), thus allowing for sufficient mixing of material from the innermost region with He-shell material. *In conclusion, the isotopic compositions of X grains are in better agreement with the large-scale mixing scenario based on existing stellar nucleosynthesis models and observations for CCSNe.*

In the following, we will, therefore, focus on the large-scale mixing scenario in which X grains sampled materials from the Fe/Ni zone (Lin et al. 2010; Liu et al. 2024a, 2024b), the inner Si/S zone (Liu et al. 2018a, 2023a), and the He (and possibly H) shells (Lodders & Fegley 1995; Nittler et al. 1996; Fedkin et al. 2010) but without any contributions from the Si/O, O/Ne, and C/O layers inbetween, as illustrated in Fig. 3. Finally, before moving onto the next section, we would like to stress that future investigations are required to explore whether $\alpha$-capture in the outer region of a CCSN can lead to significantly enhanced Ti/Si and Ti/Ca ratios as required by the X and $Si_3N_4$ grain data. This is because the models of Pignatari et al. (2015) did not include the main physical mechanisms that could shape the He abundance profile in C/O zone in the pre-explosive O/C zone, for example, secular instabilities such as rotational mixing to gravity waves (see discussion in Pignatari et al. 2013a), which could affect the effective extension and $\alpha$-capture efficiency of the C/Si zone.

### 2.1.3.3. Varying initial $^{44}$Ti/$^{48}$Ti ratios of X grains

Given that He- and H-shell materials lack $^{44}$Ti but contain $^{48}$Ti from the initial stellar composition, the proportion of He- and H-shell materials in the supernova ejecta from which X grains formed inversely affects the $^{44}$Ti/$^{48}$Ti ratio (Lin et al. 2010): a higher percentage of He- and H-shell ($^{29}$Si-rich) materials corresponds to a lower $^{44}$Ti/$^{48}$Ti ratio and a higher $\delta^{29}$Si value in the mixed supernova ejecta. In addition, the inner ($^{28}$Si-rich) Si/S zone exhibits $^{44}$Ti/$^{48}$Ti ratios that are more than an order of magnitude lower than the Fe/Ni zone (Fig. 6), which implies that a higher percentage of inner Si/S material (with respect to Fe/Ni material) in the mixed ejecta leads to reduction in both $^{44}$Ti/$^{48}$Ti and δ$^{29}$Si in the mixed supernova ejecta. Consequently, the absence of inferred initial presence of $^{44}$Ti in an X grain may be attributed to various factors: (*i*) lower-than-predicted $^{44}$Ti/$^{48}$Ti ratios produced by α-rich freeze outs in the deepest supernova regions, (*ii*) substantial dilution of Fe/Ni material with He- and H-shell material and/or inner Si/S material, with the effect of the Fe/Ni and Si/S material being more pronounced in $^{28}$Si-rich X grains and that of the He- and H-shell material in $^{29}$Si-rich X grains, and/or (*iii*) significant terrestrial Ca contamination, resulting in an undetectable $^{44}$Ca excess from the grain.

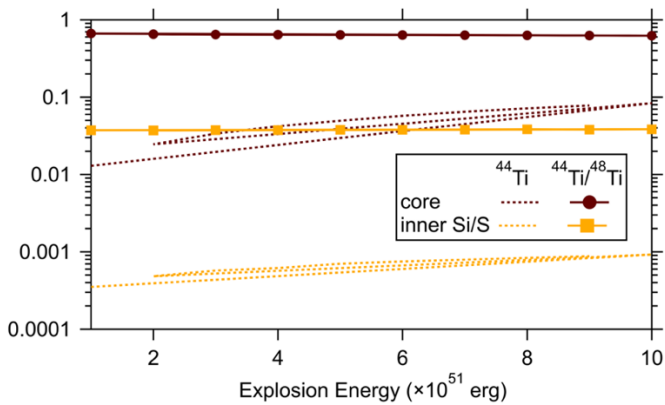

**Figure 6.** *The $^{44}$Ti/$^{48}$Ti ratio and $^{44}$Ti abundance in the Fe/Ni and inner Si/S zones predicted by the Rauscher-Meyer model calculations for a 25 $M_\odot$ progenitor massive star as a function of explosion energy. The corresponding model calculations at two of the energies (1×10$^{51}$ erg and 10×10$^{51}$ erg) are shown in Fig. 5. The Fe/Ni and inner Si/S zones are shown here for illustration of the large-scale mixing scenario that explains the isotopic compositions of X grains (see Fig. 3; Lin et al. 2010; Liu et al. 2023a, 2024b).*

We examine the three possibilities based on the X grains that exhibit significant $^{44}$Ca excesses ($\gtrsim$ 1000 ‰) shown in Fig. 4. Their large excesses minimize the effects of analytical uncertainties and potential GCE-influenced initial composition[4] on the derivation of initial $^{44}$Ti/$^{48}$Ti ratios. These X grains collectively display a compelling negative trend in Fig. 4b, despite being analyzed across different laboratories under varying conditions and using different ion microprobe techniques (including Cameca 3f-, 6f-, NanoSIMS ion microprobes). The trend in Fig. 4b points to the coproduction of $^{44}$Ti (relative to $^{48}$Ti) and $^{28}$Si (relative to $^{29}$Si), in line with the theoretical expectation for nucleosynthesis in deep supernova regions as discussed earlier. The imperfect correlation between $^{44}Ti/^{48}Ti$ and $\delta^{29}Si$ does not necessarily point to production of variable $^{44}Ti/^{48}Ti$ ratios because some small variability in the isotopic and/or elemental compositions of the two endmembers inferred from the linear mixing line – the inner $^{28}Si$-rich endmember from the Fe/Ni and Si/S region, and the outer $^{29,30}Si$-rich endmember from the He and H shells – could all contribute to the scatter in the X grain data.

---

[4] The GCE effect could lead to temporal and spatial $^{44}$Ti and $^{44}$Ca variations in the ISM and thus in the initial stellar composition (see Section 2.1.3.4. for discussion in detail). In this case, the $^{44}$Ca excess detected in an X grain would be irrelevant to the $^{44}$Ti production in its parent supernova.

In contrast to the subset of X grains (~35%) exhibiting significant $^{44}Ca$ excesses shown in Fig. 4b, the majority of X grains (~65%) reported in the literature have no or much smaller $^{44}Ca$ excesses (Stephan et al. 2024). If all X grains originated from the same supernova ejecta, then we could hypothesize that the large $^{44}Ca$ enrichments of the X grains in Fig. 4b are purely caused by their relatively low Ca/Ti ratios, and vice versa, i.e., the Scenario (*iii*) above. This hypothesis is further scrutinized based on X grains for which elemental concentration data are available (Fig. 4c). The influence of variable mixing ratios between the two endmembers can be corrected by calculating the value of $^{44}Ca/^{40}Ca_{predicted}$/solar. The first step of this calculation is based on the established correlation in Fig. 4b (solid line), which shows the $^{44}Ti/^{48}Ti$ ratio for a given $\delta^{29}Si$ value. From the predicted $^{44}Ti/^{48}Ti$ ratio, the second step in the calculation is to use the Ca and Ti concentration data to predict the $^{44}Ca$ excess for a grain, i.e., $^{44}Ca/^{40}Ca_{predicted}$/solar. This method implicitly assumes that the nonradiogenic $\delta^{44}Ca$ of an X grain equates to zero, a basis for the derivation of most initial $^{44}Ti/^{48}Ti$ data in Fig. 4b. Estimations for the lower and upper boundaries of $\delta^{44}Ca_{predicted}$ are anchored to the dashed lines shown in Fig. 4b. The resultant trend in Fig. 4c shows that the measured $^{44}Ca$ excesses in the majority of X grains fall beneath the values that we would expect if the X grains in Fig. 4c were sampled from the same supernova ejecta as those in Fig. 4b. Although potential Ti contamination, if sampled during the analyses, could lead to inflated $^{44}Ca/^{40}Ca_{predicted}$/solar values, NanoSIMS analysis of X grains at high-spatial resolution suggests that Ti contamination is rare (Liu et al. 2023a).

The scenario (*i*), i.e., reduced $^{44}Ti$ production in the deep regions of their parent supernovae, can be further excluded, given the invariable $^{44}Ti/^{48}Ti$ production ratio as a function of varying explosion energies shown in Fig. 6. The lack of any energy dependence is caused by the fact that the productions of $^{44}Ti$ and $^{48}Ti$ both increase with increasing explosion energy in the deep region of a Type II supernova. This is because $^{44}Ti$ is an $\alpha$ nuclide and $^{48}Ti$ is copiously produced as the $\alpha$ nuclide $^{48}Cr$, which subsequently decays to $^{48}V$ ($t_{1/2} = 22$ h) and then to the stable isotope $^{48}Ti$ ($t_{1/2} = 16$ d).

*By excluding the Scenarios (i) and (iii), we thus infer that the two sets of X grains with large and small $^{44}$Ca excesses likely sampled varying amounts of materials from the Fe/Ni zone ($^{28}$Si,$^{44}$Ti-rich), the inner Si/S zone ($^{28}$Si-rich, $^{44}$Ti-poor), and the outer shells ($^{29}$Si-rich, $^{44}$Ti-free), i.e., the Scenario (ii).* Variation in the relative mixing ratios between these zones explains the varying initial $^{44}$Ti/$^{48}$Ti ratios observed among X grains. In other words, the X grains with small $^{44}$Ca excesses sampled more materials from both the Si/S zone and outer shells (i.e., accounting for their comparable $^{28}$Si excesses) than the X grains with large $^{44}$Ca excesses. *The absence of inferred initial presence of $^{44}$Ti in the majority of X grains thus cannot be used as evidence for the idea that $^{44}$Ti is predominantly produced in rare supernova events.*

2.1.3.4. Potential effect of heterogeneities in the ISM on $^{44}$Ca

Two pieces of indirect evidence from presolar grain studies may still suggest that $^{44}$Ti is predominantly produced in rare supernova events, based on which a heterogenous distribution of $^{44}$Ti (and thus its decay product $^{44}$Ca) is expected (spatially and temporally) in the ISM: (*i*) the broader-than-anticipated scatter in the $^{44}$Ca/$^{40}$Ca ratios of presolar hibonite grains (Nittler et al. 2008) and (*ii*) the lack of observable correlation between the $^{44}$Ca excesses of presolar silicates and their Ti-rich subgrains (Nittler et al. 2023; Leitner et al. 2024). Nittler et al. (2008) observed varying $^{44}$Ca/$^{40}$Ca ratios in presolar hibonite grains of AGB stellar origin with similar $^{18}$O/$^{16}$O ratios, which infer a similarity in the initial stellar metallicities of their parent stars (Nittler 2009). Since AGB nucleosynthesis barely produces any $^{44}$Ca, this observation suggests that AGB stars that were born with similar metallicities, i.e., with similar $^{18}$O/$^{16}$O ratios, had relatively variable initial $^{44}$Ca/$^{40}$Ca ratios, thus corroborating the expected heterogenous distribution of $^{44}$Ti (and thus its decay product $^{44}$Ca) in the ISM due to the dominant production of $^{44}$Ti in rare supernova events. In addition, Nittler et al. (2023) and Leitner et al. (2024) observed that in two presolar silicate grains their $^{44}$Ca excesses were not spatially correlated with $^{48}$Ti signals. One of the silicate examples is shown in Fig. 7. In Fig. 7, we see that the small $^{44}$Ca excess of the silicate grain remains constant with varying $^{48}$Ti/$^{40}$Ca ratios, which is in great contrast to the linear correlation observed in the X grain shown in Fig. 2e and thus points to a nonradiogenic origin for the observed small $^{44}$Ca excess. Together with the hibonite grain data from Nittler et al. (2008), the presolar silicate data provide further evidence for the heterogenous distributions of $^{44}$Ti and thus $^{44}$Ca in the ISM. Furthermore, the grain data in Fig. 7 deviate from the homogenous GCE trend reported by Nittler et al. (2008) based on the GCE calculation of Timmes et al. (1995), and its $^{44}$Ca excess suggests local $^{44}$Ca enrichments in the ISM, which could have resulted from rare supernova events. A caveat is that the evidence abovementioned directly reflects the effect of heterogeneous ISM composition on the $^{44}$Ca/$^{40}$Ca ratio, which could have been caused by many different astrophysical processes, e.g., insufficient mixing of supernova ejecta with the ISM material (e.g., Lugaro et al. 1999; Nittler 2005). *In conclusion, presolar grain data provide no direct support for the suggestion of predominant $^{44}$Ti production in rare types of supernova events.*

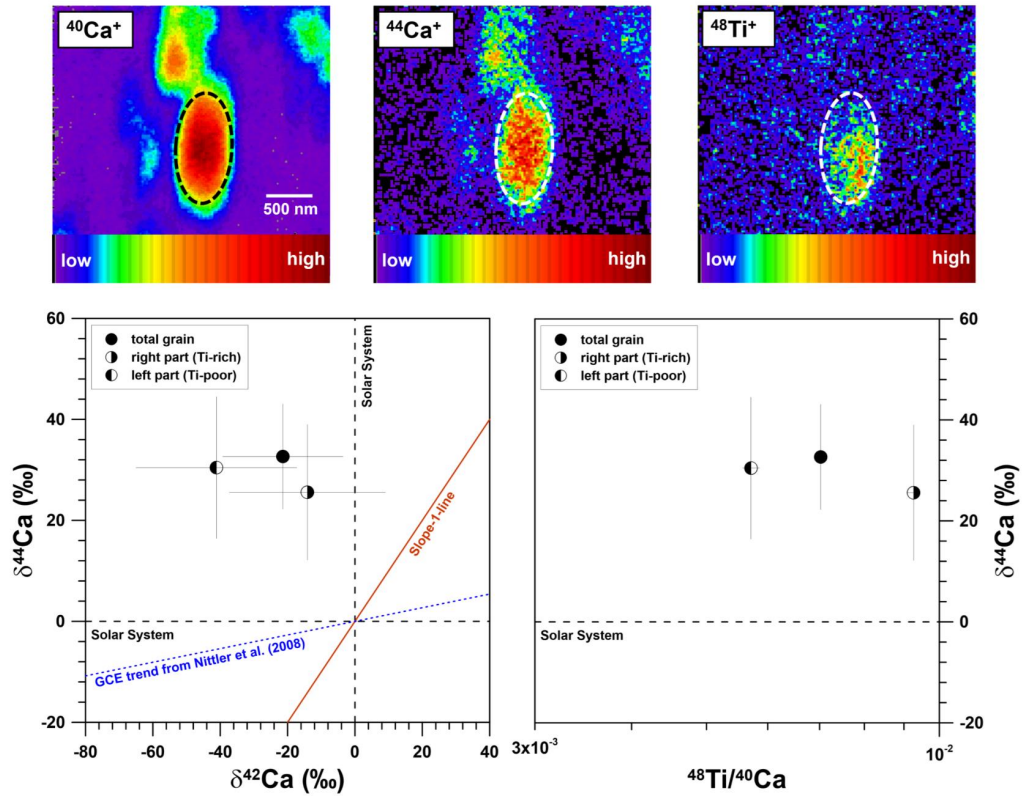

**Figure 7.** *NanoSIMS Ca-Ti isotope data of a $^{17}$O-rich silicate grain of low-mass AGB stellar origin from Leitner et al. (2024). The upper panel shows $^{40}$Ca, $^{44}$Ca, and $^{48}$Ti ion images (with color bars shown at the bottom) collected from the silicate grain (denoted by dashed ellipses) at a spatial resolution of ~100 nm. The white scalebar in the $^{40}$Ca image denotes 500 nm. Three different ROIs were defined for reducing the Ca-Ti isotope data shown in the upper panel, as denoted by the three data points shown in the lower panel.*

## 2.2. Further Insights Into Supernova Nucleosynthesis

As discussed so far, presolar supernova grains are direct condensates from supernova ejecta and thus have recorded their isotopic compositions, offering a direct window into the intricacies of supernova nucleosynthesis. Advanced isotopic analyses harnessing both modern NanoSIMS and resonance ionization mass spectrometric (RIMS) techniques are now able to extract accurate isotope ratios across elements from C to Nd in micron-sized presolar grains at percent-level precision (Liu et al. 2022a). These multielement isotope datasets are instrumental to shed new light on nucleosynthesis processes, particularly in Type II supernovae (Hoppe et al. 2018, 2023, 2024; Kodolányi et al. 2018; Liu et al. 2018a, 2024a; Pignatari et al. 2018; Stephan et al. 2018; Ott et al. 2019), given the dominance of their dust in the presolar grain inventory. The reader is referred to the abovementioned literature for recent advancements in this field. In this section, we will primarily focus on the production of neutron-rich isotopes in supernovae and their signatures in supernova grains, setting a foundational context for the subsequent discussion on nucleosynthetic isotope variations in bulk meteorites in Section 2.3.

### 2.2.1. Neutron Burst in Type II Supernovae

Multiple lines of evidence suggest that the *r*-process does not occur in typical Type II supernovae, nevertheless, these stellar events are suggested to be the sites for other neutron-capture processes, including the weak slow neutron capture process (*s*-process) during the presupernova stellar evolution (Pignatari et al. 2010) and the neutron-burst process during the explosion (Meyer et al. 2000). Specifically, the weak *s*-process is believed to occur during the hydrostatic convective core He-burning phase and the subsequent convective shell C-burning phase of a presupernova massive star. In contrast, the neutron-burst process takes place during a Type II supernova explosion in the He/C zone (Fig. 3). These two neutron-capture processes are similar in that they are both powered by neutrons released by the $^{22}$Ne($\alpha,n$)$^{25}$Mg reaction but at different neutron densities and on different timescales.

The degree to which ejecta carrying the imprint of weak *s*-process nucleosynthesis are incorporated into supernova grains – and the extent of any superimposed explosive nucleosynthesis signatures – is not yet fully explored. In contrast, the He/C zone, where the neutron burst occurs, is the only C-rich region across a Type II supernova. Therefore, material from the He/C zone is necessary for creating the C-rich condition to form C-rich dust. Corroborating this, neutron-burst isotopic fingerprints were discerned in presolar Type-X SiC grains (Pellin et al. 2006; Stephan et al. 2018), strengthening the argument that C-rich conditions are required for the formation of SiC (Lodders & Fegley 1995). An alternative hypothesis by Bliss et al. (2018) postulated that the neutrino-driven winds following a CCSN explosion could create a neutron-rich environment that is suitable to produce the Mo and Ru isotopic patterns observed in X grains. However, Bliss et al. (2018) calculated the neutrino-driven-wind condition based on an analytical model and focused solely on heavy-element isotopic signatures, thus making it challenging to explore the collateral consequences for the other isotopic signatures (e.g., C, Ti) of X grains. In contrast, CCSN stellar nucleosynthesis models (Woosley & Heger 2007; Heger & Woosley 2010) predict the neutron-burst process to occur within the He/C zone, and this satisfies the prerequisite of C-rich conditions for the formation of X grains. Therefore, the discussion below will focus on the neutron-burst scenario to interpret the neutron-rich isotopic compositions of X grains.

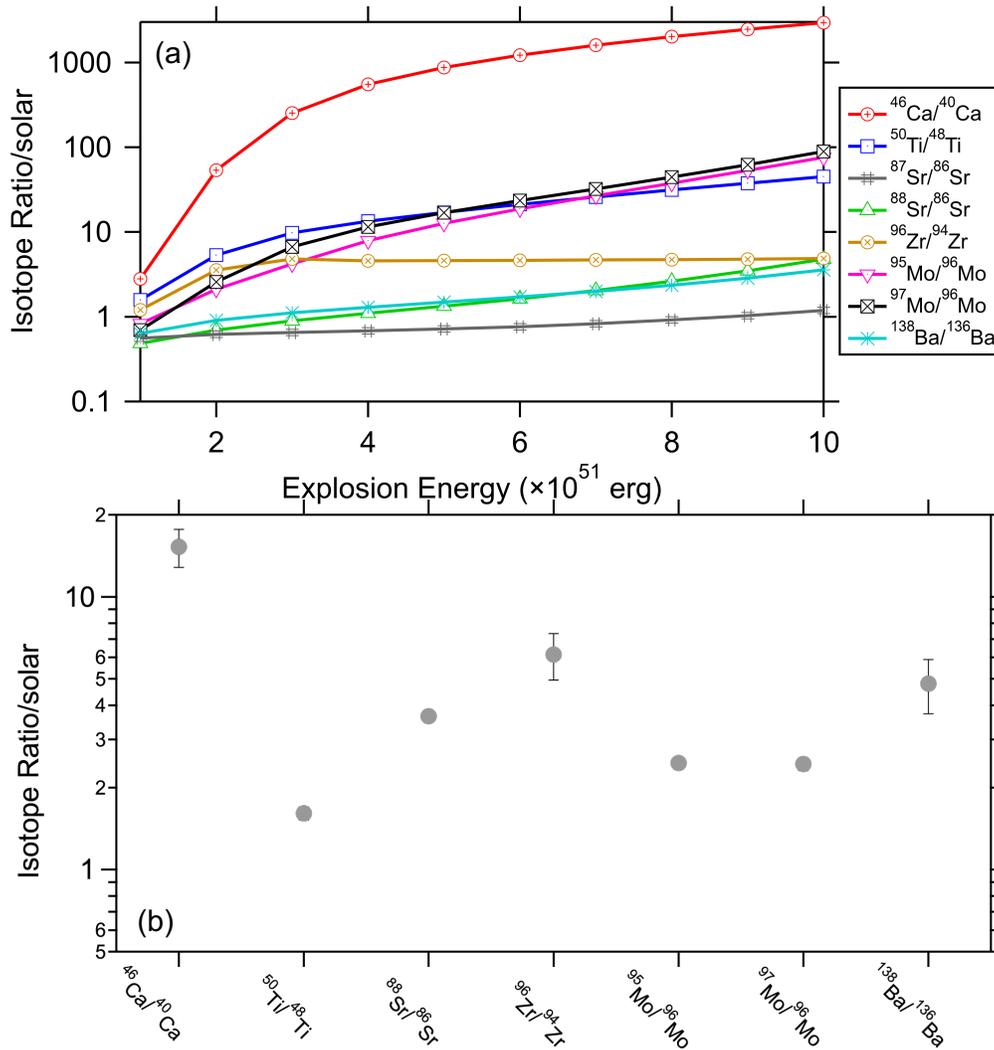

**Figure 8**. *Panel (a) displays the dependences of varying isotope ratios on explosion energy in the He/C zone predicted by the same Rauscher-Meyer models from Fig. 5. The radioactive isotopes involved in the production of the isotopes of interest all have half-lives less than tens of days and should have fully decayed to their daughter isotopes by the time of X grain condensation in the*

*supernova ejecta (Liu et al. 2018a; Ott et al. 2019; Niculescu-Duvaz et al. 2022). The isotope ratios are normalized to solar composition, i.e., unity in y-axis. Panel (b) summarizes the maximum excesses in the neutron-rich isotopes of Ca, Ti, Sr, Zr, Mo, and Ba observed in X grains so far (Amari et al. 2003; Pellin et al. 2006; Liu et al. 2024b). Plotted are 1σ errors.*

The neutron-burst process, characterized by a neutron density up to $10^{17-20}$ neutrons/cm$^3$ at temperatures around $10^9$ K (Meyer et al. 2000), predominantly produces intermediate-mass Mo isotopes like $^{95}$Mo and $^{97}$Mo (Fig. 8), and, therefore, differentiates itself from both the *s*- ($\sim 10^7 - 10^{10}$ neutrons/cm$^3$) and *r*-processes (>$10^{20}$ neutrons/cm$^3$) (Cowan & Thielemann 2004; Liu et al. 2019). Such $^{95,97}$Mo-enhanced isotopic patterns, accompanied by excesses of $^{88}$Sr, $^{96}$Zr, and $^{138}$Ba (Fig. 8b), were detected in X grains (Pellin 2000, 2006; Stephan et al. 2018), in line with neutron-burst nucleosynthesis model predictions (Meyer et al. 2000). The preservation of neutron-burst isotopic imprints for heavy elements in X grains can be ascribed to the following two aspects. (*i*) The production of these isotopes is orders of magnitude lower in the Fe/Ni and Si/S zones than in the He shell, minimizing the influence of inner supernova materials on the isotope ratios of an X grain for these elements (Stephan et al. 2018). And (*ii*) nucleosynthesis in the H envelope barely alters the initial stellar composition, which is inferred to lie close to the solar composition in the parent supernovae of X grains based on their $^{46,47}$Ti/$^{48}$Ti ratios (Liu et al. 2024b); consequently, mixing with the H envelope reduces isotope variations produced by neutron bursts but barely alters the neutron-burst isotopic patterns. Discussions on X grain data in the context of Type II supernova models that delve into these isotope ratios are reported in the literature (Pignatari et al. 2018; Stephan et al. 2018; Ott et al. 2019). Going forward, below we investigate whether the neutron-burst isotopic compositions of X grains for different elements can be explained coherently. X grains are divided into several subtypes based on Si isotopes (Stephan et al. 2024). Below, we will mention the subtypes X1 and X2 in the discussion: X1 grains are the dominant subtype of X grains and lie along the slope-2/3 line in the 3-Si-isotope plot, while X2 grains are much lower in abundance and more enriched in $^{30}$Si than X1 grains (Lin et al. 2010; Stephan et al. 2024).

Figure 8a showcases that the neutron-burst process results in the overproduction of isotopes, such as $^{46}$Ca, $^{50}$Ti, $^{88}$Sr, $^{96}$Zr, $^{95}$Mo, $^{97}$Mo, and $^{138}$Ba, with the degree of overproduction generally increasing with increasing energy, except for $^{96}$Zr. These predicted variations generally align with observations from X grains: (*i*) The comparable $^{95,97}$Mo/$^{96}$Mo ratios predicted across different energies, explain the similar levels of $^{95}$Mo and $^{97}$Mo enrichments in X grains (Fig. 8b; Pellin et al. 2000), despite that they could have originated from diverse parent supernovae whose He/C zones were heated to varying temperatures during the explosions; and (*ii*) the prominent $^{86}$Sr enhancements (suggested by the large, negative $\delta^{87,88}$Sr values that are normalized to $^{86}$Sr) observed in two X2 grains (Stephan et al. 2018) well correspond to neutron bursts at lower explosion energies (e.g., $\delta^{87,88}$Sr $\approx$ –500 ‰ at $1\times10^{51}$ erg) as compared to X1 grains (Pellin et al. 2000; Stephan et al. 2018).

However, a challenge arises in reconciling these model predictions with observed isotope data in a quantitative way. The models suggest that the neutron-burst process should yield $^{95,97}$Mo excesses at levels an order of magnitude greater than those of $^{88}$Sr and $^{138}$Ba and at least as much as $^{96}$Zr (Fig. 8a). Yet, this expected excess pattern has not been identified in the multielement isotopic composition of any known X grains (Fig. 8b; Pellin et al. 2000): the maximum excesses in $^{95,97}$Mo observed in X grains so far are not higher than those in $^{88}$Sr, $^{96}$Zr, and $^{138}$Ba. A possible explanation for the data-model discrepancy might be asteroidal and/or terrestrial Mo contamination, which can lower the observed excesses in $^{95,97}$Mo without affecting other isotope ratios (Liu et al. 2019, 2022c). Under this premise, the upper limits of observed enrichments in $^{88}$Sr, $^{96}$Zr, and $^{138}$Ba, of approximately 2000 – 3000 ‰ (Fig. 8b; Pellin et al. 2000; Stephan et al. 2018), could correspond to neutron-burst processes occurring at $(9–10) \times 10^{51}$ erg in the 25 $M_\odot$ supernova model shown in Fig. 6. Compared to astronomical observations of supernova remnants, which suggest a mean energy per unit mass of $0.85 \times 10^{51}$ erg /10 $M_\odot$ (Rubin et al. 2016), the inferred explosion energies of up to $(9–10) \times 10^{51}$ erg / 25 $M_\odot$ from the presolar grain data are significantly higher. However, these values still fall within the range observed among Type II supernova remnants $(0.2–20) \times 10^{51}$ erg /10 $M_\odot$ and may be related to explosion asymmetries or progenitor massive stars with lower initial stellar masses. On the other hand, the observed data-model discrepancies could imply uncertainties in current neutron-burst model calculations for the following reasons. The model calculations suffer from uncertainties in the adopted nuclear reaction rates, especially given that the neutron-capture and beta-decay rates of many short-lived isotopes involved in the neutron-capture path are based on theoretical calculations. Recently, efforts on providing experimental constraints have been pursued (e.g., Spyrou et al. 2024), making it promising to have more reliable neutron-burst model predictions in the foreseen future. In addition, the accuracy of neutron-burst model calculations is also affected by the adopted isotope abundances produced by presupernova nucleosynthesis, and neutron-capture processes in presupernova massive stars need to be fully explored.

Another problem is that at such high energies the neutron-burst nucleosynthesis is predicted to produce substantial excesses in $^{46}Ca$ ($2 \times 10^6$ ‰) and $^{50}Ti$ ($4 \times 10^4$ ‰). Yet, the observed $^{50}Ti$ excesses in X grains peak at roughly 700 ‰, which is substantially lower than expected, with many grains showing smaller excesses (Fig. 8b; Liu et al. 2024b). While $^{46}Ca/^{40}Ca$ ratios for X grains remain unknown due to their general lack of intrinsic Ca, recent analyses of Murchison supernova $Si_3N_4$ grains – isotopic twins of X grains (Nittler et al. 1995; Lin et al. 2010) – have uncovered $^{46}Ca$ excesses up to 10,000‰ (Fig. 8b; Liu et al. 2024b). These observed excesses in both $^{46}Ca$ and $^{50}Ti$ (Fig. 8b) fall nearly two orders of magnitude below the model predictions. As discussed earlier, this discrepancy can be reconciled by incorporating material from the Si/S and/or Fe/Ni zones, where significant amounts of $^{40}Ca$ and $^{48}Ti$ are produced.

*In conclusion, the isotopic compositions of Sr, Zr, and Ba in X grains can be explained by the neutron-burst process occurring, for example, at explosion energies up to (9–10) × $10^{51}$ erg in a 25 $M_\odot$ supernova. The observed lower $^{46}Ca$ and $^{50}Ti$ excesses in supernova grains are primarily due to the incorporation of Fe/Ni and/or Si/S material, characterized by significant production of $^{40}Ca$ and $^{48}Ti$, resulting in smaller $^{46}Ca/^{40}Ca$ and $^{50}Ti/^{48}Ti$ ratios in the mixed ejecta. The lower-than-predicted levels of $^{95,97}Mo$ excesses in X grains could suggest the presence of significant Mo contamination sampled in the study of Pellin et al. (2000).*

To test the contamination hypothesis, further Mo isotope analyses using new generation RIMS instruments at higher spatial resolution (Stephan et al. 2016; Pal et al. 2022) are needed, which will help minimize the effect of potential extrinsic contamination. Moreover, uncertainties in neutron-burst model calculations also need to be examined in the future by adopting experimentally constrained reaction rates and by considering uncertainties in the isotope abundances that are produced by presupernova nucleosynthesis. Future modeling work also needs to extend the initial mass and metallicity ranges, especially given that the progenitor stars of the vast majority of CCSNe are expected to have initial masses lower than 25 $M_\odot$ according to the initial stellar mass function. Given the complex dependencies of isotope ratios on various parameters in supernova nucleosynthesis models, it is crucial to determine the properties of the parent supernovae of X grains to refine the parameter space. This refined understanding will enable us to better constrain the role of Type II supernovae in accounting for nucleosynthetic isotopic variations observed in bulk meteorites, given the confirmed contribution of the parent Type II supernovae of X grains as a source for these bulk meteoritic isotopic variations.

### 2.2.2. Explosive H Burning in Type II Supernovae

#### 2.2.2.1 Effect on neutron bursts

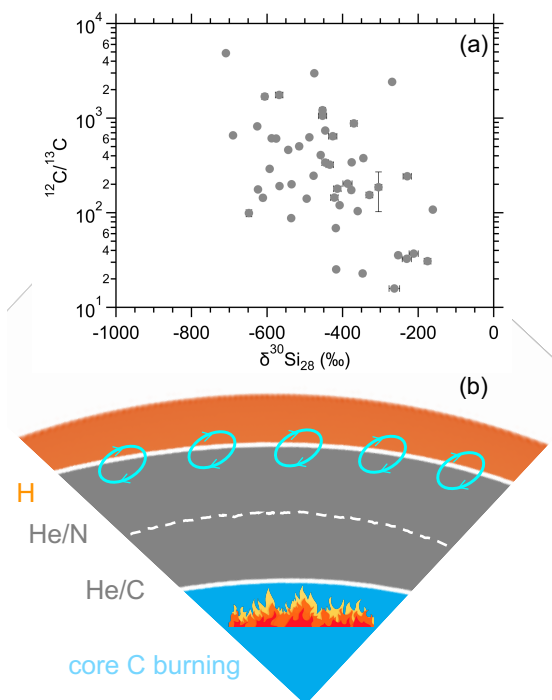

**Figure 9**. *Panel (a) presents literature C and Si isotope data for X grains from Hoppe et al. (2018) and Liu et al. (2024a). In panel (b), two possible mechanisms for introducing H into the He shell (the grey region) are illustrated. (i) Mixing of H from the bottom of the envelope into the upper layers of the He shell may occur through overshoot enabled by the convective motions of eddies in the envelope. This process results in an exponentially decaying distribution of H within the He/N zone (Liu et al., 2018b). (ii) Alternatively, mixing can be driven by the rapid release of a significant amount of energy during core C-burning, leading to an unstable convective He shell above. The instability may enable mixing of H across the entire He shell at levels reaching up to one percent (Pignatari et al. 2015).*

In this section, we will be focusing on the process of explosive H burning during Type II supernova explosions. This phenomenon specifically refers to the presence of H in the He shell during the explosions, which results in efficient production of proton-rich isotopes such as $^{13}C$, $^{15}N$, $^{25}Mg$, and $^{26}Al$ in the He shell (Pignatari et al. 2015; Schofield et al. 2022). In the absence of H, the He shell is predicted to consist predominantly of $^{12}C$ produced by triple-$\alpha$ reaction (Woosley & Heger 2007). The observed positive correlation between $^{13}C$ and $^{30}Si$ enrichments among X grains (Fig. 9a) suggests that the $^{30}Si$-rich endmember ($\delta^{30}Si \geq 0‰$, representing material from the He and H shells) has $^{12}C/^{13}C$ ratios $\lesssim 100$, which could result from the efficient production of $^{13}C$ (*i*) in the He/N zone and H envelope by presupernova nucleosynthesis and/or (*ii*) in the He/C and He/N zones by explosive H burning during the explosion. The possibility (*ii*) is favored by the fact that X grains with subsolar $^{12}C/^{13}C$ ratios exhibit a positive correlation between their $^{13}C$ and $^{15}N$ excesses (Liu et al. 2024a), which are in line with the signatures of explosive H burning in the He shell. In comparison, stellar temperatures during presupernova evolution and in the H shell during the explosion are both too low to enable hot CNO cycle for the production of $^{15}$N (Lin et al. 2010; Hoppe et al. 2018).

The exact region within the He shell where explosive H burning takes place is still debated due to the still unknown mechanism by which H is introduced into the He shell. Although the two processes illustrated in Fig. 9b differ in the quantity and the distribution of H within the He shell, modeling outcomes from both predict remarkably similar isotopic imprints (Pignatari et al. 2015; Liu et al. 2018b), and both provide a satisfying explanation for the C and N isotope ratios of X grains (Fig. 9a).

The two mechanisms, however, diverge in their impact on the neutron-burst process: H added via overshooting into the He/N zone barely influences the neutron-burst process in the He/C zone, whereas the presence of H in the He/C zone can suppress the neutron source for the neutron-burst process because $^{22}$Ne more readily captures protons via $^{22}$Ne$(p,\gamma)^{23}$Na rather than $\alpha$ particles via $^{22}$Ne$(\alpha,n)^{25}$Mg (Pignatari et al. 2015). In comparison, X grain data studied so far mostly show $^{49,50}$Ti excesses (Liu et al. 2018a, 2023a, 2024b), which can be explained by the neutron-burst process in the He/C zone, in addition to $^{49}$V decay (to $^{49}$Ti with $t_{1/2} = 330$ d) in the inner Si/S zone (Liu et al. 2018a, 2023a). The common presence of neutron-burst isotopic signatures in the parent supernova ejecta of X grains, is further supported by the observation of large $^{46}$Ca excesses, which are distinctive isotopic signatures of the neutron-burst process, in all four Si$_3$N$_4$ grains studied by Liu et al. (2024b). The Ti and Ca isotope data for SiC X and Si$_3$N$_4$ grains, respectively, were mostly collected with NanoSIMS instruments in imaging mode at high spatial resolution (~100 nm, e.g., Fig. 2), based on which extrinsic Ti and Ca contamination would have been recognized and suppressed if existed. In addition, presolar SiC grains are rich in Ti and Si$_3$N$_4$ in Ca (Liu et al. 2024b) so that the effects of extrinsic contamination were mostly small. In contrast, heavy-element isotope data for SiC X grains were collected with RIMS instruments in spot mode at 1−5 μm (Pellin et al. 2000, 2006; Stephan et al. 2018), and it was shown that the RIMS data for presolar SiC grains were affected by Sr, Ba, and Mo contamination (Barzyk et al. 2007; Liu et al. 2022c). Thus, the lack of neutron-burst isotopic signatures in some of the X grains studied by Pellin et al. (2000, 2006) is possibly caused by contamination and thus do not argue directly against the common presence of neutron-burst isotopic signatures in X grains. *The suggested common presence of neutron-burst isotopic patterns in X and Si$_3$N$_4$ grains could rule out the mixing scenario on the basis of core C-burning, although it has been argued that non-uniform, localized* *mixing of H induced by core C-burning could permit neutron bursts in less affected regions of the He/C zone (Hoppe et al. 2018; Pignatari et al. 2018).*

In general, the deactivation of neutron burst in the He/C zone significantly reduces the production of certain neutron-rich isotopes like $^{60}$Fe (e.g., Limongi & Chieffi 2006). Future studies need to investigate the effect of H ingestion on affecting the $^{60}$Fe production based on the multiple pieces of astronomical and geological evidence for the amounts of fresh $^{60}$Fe produced from nearby Type II supernovae (Binns et al. 2016; Fimiani et al. 2016; Wallner et al. 2016).

## 2.2.2.2 Effect on inferring stellar origins for presolar grains

Independent of the mechanism for inducing explosive H burning into the He shell, the probable occurrence of this process in the parent supernovae of X grains initiated a reevaluation of the stellar origins of several other types of presolar grains. This includes grains previously attributed to nova origins (putative nova grains hereafter) and Type AB grains, both of which exhibit large $^{13}$C enrichments but different ranges of $^{14}$N/$^{15}$N ratios and distinct Si isotopic compositions (Pignatari et al. 2015; Liu et al. 2016, 2017, 2018b; Hoppe et al. 2019), as well as $^{25}$Mg-rich Group 1 presolar silicates that are characterized by $^{17}$O enhancements (Leitner & Hoppe 2019). The detected enrichments of *p*-process isotopes in one such grain (Savina et al. 2007) supports a potential supernova provenance for AB grains as well as the initial $^{44}$Ti presence in another grain (Liu et al. 2023b). The hypothesized supernova origins of putative nova SiC grains and $^{25}$Mg-rich Group 1 silicates, however, should be considered cautiously. This is because their characteristic isotopic signatures – excesses in $^{13}$C, $^{15}$N, $^{17}$O, $^{25}$Mg, or $^{26}$Al – point solely to high-temperature H-burning nucleosynthesis, which is not exclusive to Type II supernovae but could occur in various stellar environments such as novae (José & Hernanz 2007), intermediate-mass AGB stars (Karakas & Lugaro 2016), and super-AGB stars (Doherty et al. 2014). Interestingly, due to its fragile nature, $^{18}$O can be easily destroyed during CNO cycle, leading to large $^{18}$O depletions that characterize explosive H burning in the He shell during a Type II supernova explosion, although these depletions could be erased if mixing occurred with adjacent He-and H- shell material and the pre-supernova material (Pignatari et al. 2015; Leitner & Hoppe 2019). In comparison, Type II supernova oxides (i.e., Group 4 grains) and graphite grains with the inferred initial presence of $^{44}$Ti are characterized by $^{18}$O enrichments (Nittler et al. 2008; Groopman et al. 2012), pointing to the occurrence of the reactions $^{14}$N$(\alpha,\gamma)^{18}$F$(\beta^{+})^{18}$O in the absence of protons in the He/C zone.

*Despite their ambiguous stellar origins, putative nova grains, Type AB grains, and $^{25}$Mg-rich Group 1 silicates provide a unique opportunity to specifically investigate high-temperature H-burning processes, and intercomparison of their isotope data between these grain types may shed new lights onto their stellar origin(s).* Existing model predictions for novae, intermediate-mass AGB stars, and super-AGB stars face a number of problems in explaining the multielement isotope data of putative nova SiC grains and $^{25}$Mg-rich Group 1 silicates (Amari et al. 2001; Nittler & Hoppe 2005; Liu et al. 2016; Leitner & Hoppe 2019). It remains to be seen whether these data-model discrepancies can be used to firmly exclude these stellar sources or reflect modeling uncertainties, e.g., missing physics due to the simplified 1D modeling approach, largely uncertain mass-loss rates for intermediate-mass AGB stars and super AGB stars.

### 2.2.3. Production of $^{54}$Cr in Supernovae

Chromium-54 isotopic variations in bulk meteorites led Dauphas et al. (2010) and Qin et al. (2011) to independently discover presolar Cr-rich spinel grains, tens to hundreds of nanometers in size, with up to thousands of permil enrichments in $^{54}$Cr. Initial assessments of Cr isotopes in nanospinel grains using NanoSIMS instruments equipped with a DuoPlasmatron O$^-$ ion source were constrained by a spatial resolution of ~0.5–1.0 μm, leading to significantly diluted isotopic anomalies due to the sampling of Cr contamination from neighboring grains (Dauphas et al. 2010; Qin et al. 2011). Subsequent advancements in the reduced beam size of the Hyperion O$^-$ plasma ion source enabled in situ Cr isotope analyses at refined resolution (~100 nm), and this led to more accurate determinations of intrinsic $^{54}$Cr enrichments of up to 50 times its solar abundance (Nittler et al. 2018).

Substantial $^{54}$Cr can be produced by (*i*) neutron-burst processes in the O/Ne, C/O, and He/C zones in Type II supernovae (den Hartogh et al. 2022) and (*ii*) quasi-equilibrium in high-density Type Ia supernovae (Woosley 1997) and (*iii*) quasi-equilibrium in ECSNe (Wanajo et al. 2013). The pronounced $^{54}$Cr enrichments in nanospinel grains, therefore, reveal inconclusive insights into their exact stellar origin. Nittler et al. (2018) suggested an ECSN or Type Ia supernova origin for these grains, but detailed comparisons with Type II nucleosynthesis models by den Hartogh et al. (2022) suggested that the nanospinel data from Nittler et al. (2018) could also be consistently explained by nucleosynthesis in the O/Ne and C/O zones of a Type II supernova, underscoring the ambiguity in determining the stellar source of the grains solely based on the Cr isotope ratios. This complexity reflects the similarities of nucleosynthesis products across the supernova types, despite their distinct evolutionary paths and explosion mechanisms. In particular, given the diversity in nucleosynthesis across various zones of Type II supernovae and the potential mixing of materials from these regions during the explosion, categorically ruling out a Type II supernova as the progenitor of certain presolar grains or as a contributor to certain meteoritic isotopic anomalies is challenging.

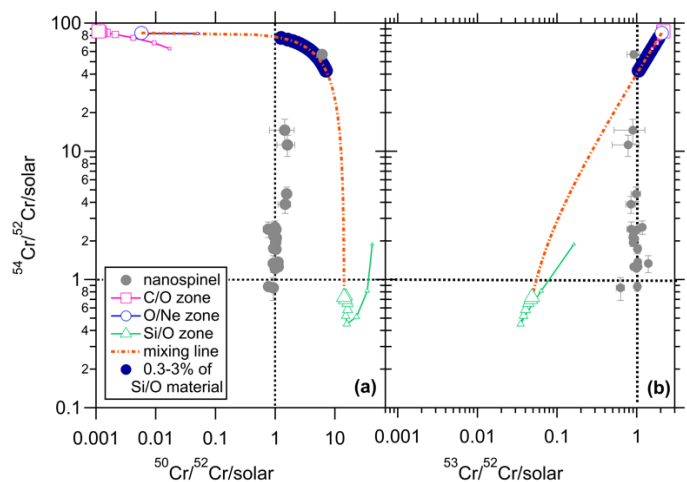

**Figure 10**. *Presolar nanospinel grain data from Nittler et al. (2018) are compared to the Rauscher-Meyer CCSN models shown in Figs. 6 and 8. The black dashed lines represent the solar ratios. Predictions for different supernova zones are shown as symbols with lines (as indicated in the legend), with the symbol size increasing with increasing explosion energy from $1 \times 10^{51}$ to $10 \times 10^{51}$ erg. The orange dashed line represents mixing of materials from the O/Ne and underlying Si/O zones at $10 \times 10^{51}$ erg, and the blue shaded region along the orange line highlights the compositions predicted by mixing 0.3−3% of material from the Si/O zone into the O/Ne zone. The decay of $^{53}Mn$ to $^{53}Cr$ was not included in the model calculations, given its relatively long half-life ($t_{1/2} = 3.7$ Ma) and unknown Mn/Cr ratios in the nanospinel grains.*

Figure 10 provides an illustrative example for Cr isotopes and reveals that supernova zones possess distinct Cr isotopic signatures, delineating that only a narrow range of mixing ratios between zones rich in $^{54}Cr$ (such as He/C, C/O, O/Ne) and those rich in $^{50}Cr$ (such as Si/O, Si/S) can approximate the most extreme Cr isotopic composition observed in nanospinel grains (as indicated by the blue shaded region in Fig. 10).[5] The models predict substantial $^{52}Cr$ production in the Fe/Ni zone via α-rich freeze outs and unaltered initial Cr isotopic composition in the He/N and H zones. The information above allows us to infer that (*i*) the isotopic compositions within the blue shaded area could also be reproduced by incorporating material from the C/O and Si/S zones, which exhibit Cr isotopic ratios akin to O/Ne and Si/O zones, respectively, and (*ii*) the less extreme

---

[5] For the data-model comparison in Fig. 10, we assumed that the ion signals at mass 50 were fully from $^{50}Cr$ and not $^{50}Ti$. The assumption is needed because the isobaric interreference at mass 50, if existed, could not be resolved in the NanoSIMS measurements of Nittler et al. (2018).

Cr isotopic compositions of the nanospinel grains could emerge from dispersing the described ejecta with contributions from the He/N zone and H envelope.

*In comparison to Type II supernova nucleosynthesis (in high-entropy conditions), the quasi-equilibrium nucleosynthesis process, occurring in low-entropy conditions predicted for ECSNe and for rare types of high-density Type Ia supernovae, preferentially produces $^{54}Cr$ with minimal impact on the other Cr isotopes (Woosley 1997; Wanajo et al. 2013; Jones et al. 2019b), thus providing a direct explanation to the observed isotopic compositions of nanospinel grains.* It is commonly believed based on spectroscopic observations that Type Ia supernova explosions produce significantly less dust than Type II supernova explosions (Gomez et al. 2012). Recent mid-infrared observations of the Type Ia supernova 2018evt, revealed the presence of about 1 $M_\odot$ of newly formed dust after the supernova explosion (Wang et al. 2024), which is comparable to the maximum amount of dust observed in Type II supernova remnants (Niculescu-Duvaz et al. 2022). Furthermore, the simulations of Jones et al. (2019b) predict efficient productions of O, Mg, Si in and large mass ejections from thermonuclear ECSNe, thus providing a preferential condition for dust formation. Information on the existence of ECSNe and their occurrence rate and dust formation efficiency based on observations, is urgently needed to evaluate their possible contribution to the presolar grain inventory in the solar system. Future detailed isotopic analyses encompassing elements such as O, Ti, Fe, and Cr with both NanoSIMS and RIMS instruments are expected to enhance our ability to discriminate among the proposed formation scenarios for these grains.

## 2.3. Supernova Nucleosynthesis and Nucleosynthetic Variability in Meteorites

### 2.3.1. Bulk Meteoritic Isotopic Variations

Nucleosynthetic isotope variations in bulk meteorites have been observed for a variety of elements, including major elements like Fe (e.g., Cook & Schönbächler 2017; Schiller et al. 2020; Hopp et al. 2022), trace elements like Mo and Pd (e.g., Ek et al. 2020), refractory elements like Ca and Ti (e.g., Dauphas et al. 2014), and moderately volatile elements like K and Zn (e.g., Martins et al. 2023; Nie et al. 2023). In addition, intra- and inter-variations have been observed in the nucleosynthetic isotope compositions of non-carbonaceous (NC) and carbonaceous (CC) meteorites, which are assumed to have formed in the inner and outer parts of the disk, respectively (Leya et al. 2008; Desch et al. 2018). The NC-CC isotopic differences may indicate distinct nucleosynthetic isotopic signatures in these reservoirs, although the exact origin of these differences is still debated. For instance, when considering Mo isotopes, the variations within NC and CC meteorites clearly point to $s$-process[6] variations (e.g., Dauphas et al. 2002; Burkhardt et al. 2011). However, the cause of inter-variations in Mo isotopic compositions between NC and CC meteorites remains ambiguous. It was proposed that the offset between the trends in NC and CC meteorites was due to an additional $r$-process component in the CC isotopic reservoir (e.g., Budde et al. 2016). However, Stephan & Davis (2021) showed that $s$-process variability could alternatively explain the observed offsets in all Mo isotopes. In this scenario, the CC isotopic reservoir sampled more $s$-process material formed at higher-than-average neutron densities than the NC isotopic reservoir, given that this material was homogenously distributed within the two reservoirs. It is unclear, however, how such two reservoirs remained separated in the ISM from the time the dust was ejected by their parent AGB star(s) to the formation of the solar system, or if they separated instead within the protoplanetary disk due to, e.g., size differences (see Section 2.3.3).

In addition, enrichments in neutron-rich isotopes like $^{48}$Ca, $^{50}$Ti, $^{54}$Cr (e.g., Niederer et al. 1980; Papanastassiou 1986; Leya et al. 2009; Chen et al. 2015; Kööp et al. 2018; Akram et al. 2013) as well as $r$-process isotopes of elements with A < 140 (e.g., Brennecka et al. 2013; Akram et al. 2013) and $p$-process isotope enrichments (e.g., Charlier et al. 2021), were identified in calcium, aluminum-rich inclusions (CAIs). These inclusions are, so far, the oldest dated material in the solar system and represent the earliest condensates in the solar nebula (MacPherson 2014). Given that bulk meteorites generally exhibit variations in the neutron-rich isotopes of refractory elements like $^{48}$Ca and $^{50}$Ti, it has been proposed that CAIs (and potentially CAI-like dust that displays isotopic similarities to CAIs but differs chemically) are largely responsible for these observed anomalies (Leya et al. 2008; Trinquier et al. 2009; Akram et al. 2015; Burkhardt et al. 2019; Yap & Tissot 2023).

*In summary, the pattern of nucleosynthetic variations in bulk meteorites suggests at least two distinct nucleosynthetic components at play: s-process nucleosynthetic products most likely carried by presolar AGB stardust grains and another unknown component with its isotopic*

---

[6]The $s$-process nucleosynthesis hereafter refers to the main $s$-process specifically, occurring in AGB stars (Lugaro et al. 2023). It is noteworthy that there is also the weak $s$-process operating in massive stars (see discussion in section 2.2.1), which can efficiently produce heavy elements before the first $s$-process peak at Sr (see Pignatari et al. 2010 and Lugaro et al. 2023 for details).

*compositions recorded by CAIs.* It is also important to note that most of the dust – from which our Sun and planets formed – had an averaged isotope composition of the local ISM and only ~3% of the ISM dust (i.e., presolar grains) kept the extreme isotopic compositions inherited from their parent stars (Hoppe et al. 2017; Ek et al. 2020). These presolar grain components are then responsible for the nucleosynthetic signatures found in bulk meteorites.

### 2.3.2. Linking Meteoritic Nucleosynthetic Variations to Astrophysical Sites
2.3.2.1. Nucleosynthetic variations in CAIs and their astrophysical sites

The stellar sites of the *r*- and *p*-processes remain a subject of ongoing debate (Arnould & Goriely 2003; Tanvir et al. 2017; Kajino et al. 2019). The proposed astrophysical sites for the *r*-process include NS-NS mergers, black hole and NS mergers, and rare types of Type II supernovae, i.e., special classes of Type II supernovae with fast rotation and high magnetic fields (Cowan & Thielemann 2004; Cowan et al. 2021), while those for the *p*-process include Type Ia and Type II supernovae (Travaglio et al. 2011; Sasaki et al. 2022; Roberti et al. 2023). In addition, a weak *r*-process, i.e., reaching up to atomic mass ≈ 100 only, could potentially occur in ECSNe (Wanajo et al. 2011), and several processes producing isotopes up to this mass may occur during α-rich freeze outs and in the neutrino winds from the nascent neutron star during CCSNe (e.g., Akram et al. 2013).

While neutron-burst processes in Type II supernovae can account for the $^{50}$Ti and $^{54}$Cr enrichments of CAIs, neutron-burst models predict orders of magnitude higher production of $^{46}$Ca than $^{48}$Ca (see discussion in Section 2.2.1), inconsistent with the Ca isotope data of CAIs that shows distinct $^{48}$Ca enrichments (Chen et al. 2015; Kööp et al. 2018). In addition, deep supernova regions like the Fe/Ni zone are characterized by prolific production of the α-isotope $^{40}$Ca with no significant production of $^{48}$Ca, i.e., $\delta^{48}$Ca = −1000‰. This is because $^{48}$Ca cannot be produced in high-entropy conditions that are characteristic of Type II supernovae (Meyer et al. 1996). Meyer et al. (1996) suggested that low-entropy conditions like those that may be achieved in rare Type Ia supernovae are ideal for efficient $^{48}$Ca production. This was later confirmed based on detailed nucleosynthesis model calculations of high-density Type Ia supernovae and ECSNe (Woosley 1997; Wanajo et al. 2013; Jones et al. 2019b). In addition, Ni and Fe isotope data of bulk meteorites suggest that both Type II and Type Ia supernova nucleosynthesis can explain the observed isotopic signatures (Nanne et al. 2019; Cook et al. 2021; Hopp et al. 2022). Moreover, it has been observed that variations in $^{46}$Ti and $^{50}$Ti, the most proton- and neutron-rich Ti isotopes, respectively, are well correlated among CAIs (Davis et al. 2018)[7]. Therefore, it appeared that multiple stellar sources, e.g., Type Ia and Type II supernovae, contributed materials to the CAI-forming region.

More recently, Meyer & Bermingham (2023) showed that isotopes produced by different nucleosynthetic processes, including the neutron-rich isotopes $^{48}$Ca and $^{96}$Zr, the proton-rich isotope $^{46}$Ti, the $p$-process isotope $^{84}$Sr, could be produced together in an exploding white dwarf, corresponding to a rare high-density Type Ia supernova or an ECSN event. This is because the required distinct nucleosynthetic processes can occur at varying temperatures and densities at different stellar radii in these events. Detailed modeling of the stellar evolution of white dwarf explosions, coupled with nucleosynthesis calculations, is needed to test the proposal of Meyer & Bermingham (2023). GCE calculations are also needed to evaluate the likelihood of such events in contributing to the observed meteoritic isotopic variations in the early solar system.

*In conclusion, existing isotopic evidence suggests that the nucleosynthetic variations of CAIs require the contribution of materials to the CAI-forming region from (i) multiple stellar sources, e.g., Type Ia and Type II supernovae or (ii) a rare high-density Type Ia supernova or an ECSN event as suggested by Meyer & Bermingham (2023).*

### 2.3.2.2. *s*-Process nucleosynthetic variations

In contrast to ambiguities in the stellar sites of the neutron-rich isotopes of the intermediate-mass elements Ca, Ti, and Cr, and the $r$- and $p$-process isotopes, the discovery of Tc absorption lines in Mira-type variable stars by astronomer Paul Merrill (Merrill 1952) provided concrete evidence that the short-lived radioactive isotope $^{99}$Tc ($t_{1/2}$ = 0.21 Ma) is freshly synthesized in stellar interiors through the $s$-process along the beta stability valley. This discovery established a direct connection between $s$-process nucleosynthesis and low-mass AGB stars. Various research groups have developed sophisticated stellar models to elucidate the evolution of AGB stars and the associated nucleosynthesis processes (Cristallo et al. 2009; Karakas & Lugaro 2016; Battino et al. 2019; Vescovi et al. 2020). These AGB stellar models confirmed that the $s$-process predominantly operates in the He-intershell regions of low-mass AGB stars (1.5 $M_\odot \lesssim M \lesssim 4\ M_\odot$) (Gallino et al. 1988). The presence of $s$-process isotopic signatures in presolar SiC and graphite grains for heavy elements like Zr and Mo provides valuable insights that have led to stringent constraints on AGB modeling parameters (Lugaro et al. 2003; Vescovi et al. 2020; Liu et al. 2018c; 2022a). With a reasonably comprehensive understanding of the *s*-process in AGB stars, Ek et al. (2020), based on nucleosynthetic isotope data of Pd, Mo, Ru and Zr from in bulk meteorites, identified presolar dust from high-metallicity AGB stars as the primary source that is responsible for the *s*-process isotopic variabilities of the bulk meteorites. Smaller isotope effects in heavier element (e.g., Nd, Hf) are predicted as a collateral consequence of high-metallicity AGB stars being sources for the *s*-process variations, and this was indeed observed in the meteorite record (e.g., Ek et al. 2020; Akram et al. 2013).

---

[7] Interestingly, variations in $^{46}$Ti and $^{50}$Ti are not correlated in chondritic matrices according to leaching experiments (Trinquier et al. 2009). This could be supporting evidence that small CAI dust originating from different stellar production sites was fused into forming the large CAIs present in chondrites.

### 2.3.3. Thermal Processing versus Size Sorting of Dust Carriers

Considering that Pd is more volatile than Mo and Ru, Ek et al. (2020) further proposed that the diminished Pd isotopic effects stem from incomplete elemental condensation around AGB stars. They also proposed that the observed *s*-process variations are due to potential destruction of more fragile ISM dust, relative to AGB dust in the inner protoplanetary disk in the early solar system. This scenario predicts the lack of nucleosynthetic variations observed in the more volatile *s*-process elements such as Cd (Toth et al. 2020). On the other hand, recent findings have revealed substantial isotopic variations in the neutron-rich isotopes of the moderately volatile elements K and Zn ($^{40}$K and $^{70}$Zn) within and between CC and NC meteorites (Martins et al. 2023; Nie et al. 2023). The discrepancy between Cd and K, Zn isotopes, highlights the role of dust carrier in controlling the preservation of its carried isotopic anomaly in meteorites. It is conceivable that $^{40}$K and $^{70}$Zn were enclosed in a thermally more robust carrier phase than, e.g., Cd, and thus their signature survived from the thermal processing. Given the still unclear nucleosynthetic origins of Zn and K isotopes combined with the unknown nature of their carriers, the observed K and Zn isotopic anomalies currently provide no direct clues to test the scenario of thermal processing.

Besides thermal processing, size sorting has been postulated as an alternative mechanism for introducing nucleosynthetic isotope variations that are assumed to be carried in diverse presolar phases within bulk meteorites (Pignatale et al. 2017, 2019). The gas-dust model simulations of Hutchison et al. (2022) used the two dust populations with their grain size distributions defined by the meteorite record: the nano-um sized presolar grain fraction and solar system silicates that grew up to mm-cm size (i.e. single grains, chondrules). Their results show that, if presolar grains remain small and do not form aggregates, they can become enriched in the outer disk through dust transport by viscous expansion within the protoplanetary disk. This enrichment mechanism may potentially explain the amplified isotopic signatures, such as those seen in $^{54}$Cr, $^{50}$Ti, and $^{96}$Zr, within the CC reservoir located at greater heliocentric distances in comparison to the NC reservoir. Moreover, nucleosynthetic effects of the presolar grain fraction can be reduced in pressure bumps, e.g, produced by Jupiter in the disk, because the larger solar system dust is trapped in the bumps, while the smaller presolar grains are coupled to the gas and can move more freely through the bump.

Finally, investigations into presolar grains suggest an interesting trend. Specifically, studies on presolar grains revealed that the abundances of presolar grains from low-metallicity AGB stars (e.g., Type-Y and -Z SiC grains) and supernovae (e.g., Type-X SiC grains, Group 3 and 4 silicates, and Group 4 oxides) both increase with decreasing grain size (Hoppe et al. 2010, 2015; Liu et al. 2022b). It is important to note that the size disparities between grains from distinct stellar sources may be subtle, with most grains falling within the submicron (and maybe even smaller) range. Low-metallicity AGB dust is expected to carry stronger $s$-process isotopic signatures for the following reasons. Stellar observations of $s$-process element enhancements at the surface of C-rich stars suggest increasing $s$-process efficiency with decreasing stellar metallicity (e.g., Lugaro et al. 2020). This is because the $s$-process efficiency is linearly correlated with the number of neutrons per Fe atom – the seed for the $s$-process. In low-metallicity AGB stars, the amount of Fe is reduced because of the lowered metallicity, and the number of neutrons remains similar. Thus, the $s$-process occurs at enhanced neutron densities in low-metallicity AGB stars (see Käppeler et al. 2011 and Liu et al. 2022a for reviews of the $s$-process in AGB stars).

To explore the effect of size disparities further, future simulations are warranted to assess the feasibility of enriching relatively smaller presolar grains from lower-metallicity AGB stars and supernovae in the outer protoplanetary disk. If such enrichment processes existed, it could lead to (*i*) higher concentrations of low-metallicity AGB dust carrying $s$-process isotopic signatures at higher-than-average neutron densities (Liu et al. 2019, 2022c), providing a mechanism for the scenario proposed by Stephan & Davis (2021) to explain the NC-CC dichotomy for Mo isotopes, and (*ii*) supernova dust hosting neutron-rich isotopic signatures like $^{54}$Cr within the outer protoplanetary disk. While these potential enrichments in low-metallicity AGB and supernova dust within the outer protoplanetary disk may offer an explanation for the observed variations in *s*-process and neutron-rich isotopic signatures among bulk meteorites, it is noteworthy that SiC grains of Type Z and Y from low-metallicity AGB stars have been observed to show heavy-element isotopic signatures that differ from low-metallicity AGB model predictions (Liu et al. 2019) and more investigations are needed.

## 3. CONLUSIONS

In conclusion, the presence of presolar supernova grains in extraterrestrial materials and their distinctive isotopic signatures have provided ground-truth information to investigate a variety of nucleosynthesis processes in supernovae, especially in Type II supernovae. Below, we summarize the key implications and conclusions drawn from our overview.

1. Type II supernovae are known for their significant dust production based on spectroscopic observations, which is further corroborated by the dominance of Type II supernova dust among presolar supernova grains. Specifically, evidence for the Type II supernova origin of Type-X SiC grains includes the initial presence of $^{44}$Ti in them, demonstrated by recent high-resolution NanoSIMS imaging data that allowed direct distinguishing between radiogenic and nonradiogenic $^{44}$Ca excesses. Our analysis of the literature Ca-Ti isotope data from X grains suggested that the varying $^{44}$Ti/$^{48}$Ti ratios observed among X grains resulted from sampling materials from the $^{44}$Ti-producing Fe/Ni and Si/S zones and outer $^{44}$Ti-free He and H shells with varying mixing ratios. In addition, nonradiogenic origins of small $^{44}$Ca excesses were observed in two presolar silicate grains likely originating from low-mass AGB stars. The detected nonradiogenic $^{44}$Ca excesses imply a spatially and/or temporally heterogenous distribution of the $^{44}$Ca/$^{40}$Ca ratio in the ISM and possibly, in general, contributed to the $^{44}$Ca/$^{40}$Ca variations observed among presolar oxides and silicates of low-mass AGB stellar origin.

2. In addition to the presence of $^{44}$Ti, the Type II supernova origin of X grains is further corroborated by their other distinct isotopic signatures such as large excesses in $^{28}$Si and $^{26}$Al. The heavy-element isotopic compositions of X grains are characterized by the signatures of neutron bursts occurring in the He/C zone during the explosions of their parent supernovae. Current supernova model calculations face challenges in simultaneously explaining the neutron-burst isotopic signatures of X grains across multiple elements. Further studies are required to explore potential terrestrial/asteroidal contamination effects on grain measurements for varying elements and uncertainties in model predictions due to, e.g., uncertain nuclear reaction rates.

3. The light-element isotopic compositions of X grains indicate explosive H burning occurring in the He shell during the explosions of their parent supernovae, thus requiring mixing of H from the upper H shell into the underlying He shell. Two proposed mechanisms include H overshooting into the underlying He/N zone and H mixing throughout the entire He shell driven by rapid energy release during core C-burning (i.e., H ingestion). Model simulations based on the two mechanisms yield similar isotopic signatures and can both explain the light-element isotopic signatures of X grains. The two mechanisms, however, differ in their effects on the neutron-burst process. While mixing of H into the H/N zone has no effect on this process, the H-ingestion process is anticipated to deactivate the neutron source for the neutron-burst process in the He/C zone, which seemingly contradicts the common presence of neutron-burst isotopic signatures in X grains. Arguments, however, have been made for the H-ingestion scenario to be in line with the X grain data that non-uniform, localized mixing of H induced by core C-burning could allow for neutron bursts in less affected regions of the He/C zone. Independent of the mechanism for inducing explosive H burning into the He shell, the probable occurrence of this process in Type II supernovae has led the proposal that presolar grains that carry explosive H burning isotopic signatures could all have come from Type II supernovae. Intercomparison of the isotope data of these potential supernova grains may shed new lights onto their ambiguous stellar origin(s).

4. Presolar nanospinel grains exhibit substantial $^{54}$Cr excesses that point to their supernova origin, although the specific supernova type remains ambiguous, with high-density Type Ia, ECSNe, or Type II supernovae as potential candidates. Further investigations are essential to understand the specific mixing process required for the observed Cr isotopic signatures based on Type II supernova nucleosynthesis, quantify the relative contributions of these distinct supernova sources to the presolar dust reservoirs in the solar system, and acquire multielement isotopic compositions of presolar nanospinel grains to better distinguish between these distinct supernova sources.

5. The presence of nucleosynthetic isotope variations in bulk meteorites implies the heterogeneous

distribution of various nucleosynthetic components that at least include *s*-process nucleosynthesis component carried in presolar AGB stardust grains and an unresolved supernova component with its isotopic compositions recorded by CAIs. Multiple supernova sources, including Type Ia supernovae, Type II supernovae, and ECSNe, have been suggested to explain the isotopic composition of the CAI component. In addition, systematic differences in the isotopic compositions of NC and CC meteorites point to a separation of the two isotopic reservoirs in the protoplanetary disk in the early solar system, based on which two mechanisms, thermal processing and size sorting, have been proposed. Future studies are required to investigate the nucleosynthetic origins of observed K and Zn isotopic anomalies with the absence of Cd isotopic variations in bulk meteorites for testing the mechanism of thermal processing, and the feasibility of enriching the relatively smaller presolar grains from low-metallicity AGB stars and supernovae in the outer protoplanetary disk based on presolar grain data.

These findings above underscore the complex interplay of stellar evolution, nucleosynthesis, and mixing, and the temporal and spatial effects of GCE on the ISM composition, highlighting the need for continued research to unravel the intricate origins of isotopic anomalies in various meteoritic components and their relations.

**Compliance with Ethical Standards**: The authors declare that they have no conflict of interest.

**Acknowledgements**: This paper was instigated at the International Space Science Institute (ISSI) workshop *"Evolution of the solar system: constraints from meteorites."* The authors would like to express their gratitude to ISSI for the travel support. The authors also would like to thank Dr. Maria Bergemann, Dr. Larry Nittler, and an anonymous referee for a careful reading of the manuscript and helpful feedback and Drs. Marco Pignatari and Carolyn Doherty for insightful discussions and inputs. This work was supported by NASA through grants 80NSSC23K1034 and 80NSSC24K0132 to N.L. and 80NSSC20K0338 to B.S.M. M.L. acknowledges the support of the Lendület Program (LP2023-10) of the Hungarian Academy of Sciences and the NKFIH excellence grant TKP2021-NKTA-64. J.L. acknowledges support by the German Research Foundation (DFG) through grant LE 3279/3-1. We thank the European Union's Horizon 2020 research and innovation program (ChETEC-INFRA -- Project no. 101008324).